\documentclass[prd,twocolumn,amsmath,amssymb,nofootinbib,floatfix,superscriptaddress]{revtex4}

\usepackage{graphicx,bm,xcolor}
\usepackage{enumerate}
\usepackage{graphicx,bm}
\usepackage[normalem]{ulem}

\makeatletter
\def\graphicscale{\twocolumn@sw{0.3}{0.4}}
\def\graphicthreescale{\twocolumn@sw{0.3}{0.4}}

\begin{document}

\title{Diverse universality classes of the topological deconfinement
  transitions \\of three-dimensional noncompact lattice Abelian Higgs
  models}

\author{Claudio Bonati} 
\affiliation{Dipartimento di Fisica dell'Universit\`a di Pisa
        and INFN Sezione di Pisa, Largo Pontecorvo 3, I-56127 Pisa, Italy}

\author{Andrea Pelissetto}
\affiliation{Dipartimento di Fisica dell'Universit\`a di Roma Sapienza
        and INFN Sezione di Roma I, I-00185 Roma, Italy}

\author{Ettore Vicari} 
\affiliation{Dipartimento di Fisica dell'Universit\`a di Pisa,
        Largo Pontecorvo 3, I-56127 Pisa, Italy}

\date{\today}

\begin{abstract}
We study the topological phase transitions occurring in
three-dimensional (3D) multicomponent lattice Abelian Higgs (LAH)
models, in which an $N$-component scalar field is minimally coupled
with a noncompact Abelian gauge field, with a global $SU(N)$
symmetry. Their phase diagram presents a high-temperature Coulomb (C)
phase, and two low-temperature molecular (M) and Higgs (H) phases,
both characterized by the spontaneous breaking of the $SU(N)$
symmetry.  The molecular-Higgs (MH) and Coulomb-Higgs (CH) transitions
are topological transitions, separating a phase with gapless gauge
modes and confined charges from a phase with gapped gauge modes and
deconfined charged excitations.  These transitions are not described
by effective Landau-Ginzburg-Wilson theories, due to the active role
of the gauge modes.  We show that the MH and CH transitions belong to
different {\em charged} universality classes.  The CH transitions are
associated with the $N$-dependent charged fixed point of the
renormalization-group (RG) flow of the 3D Abelian Higgs field theory
(AHFT). On the other hand, the universality class of the MH
transitions is independent of $N$ and coincides with that controlling
the continuous transitions of the one-component ($N=1$) LAH model.  In
particular, we verify that the gauge critical behavior always
corresponds to that observed in the 3D inverted $XY$ ($IXY$) model
(dual to the 3D $XY$ vector model with Villain action), and that the
correlations of an extended charged gauge-invariant operator (in the
Lorenz gauge, this operator corresponds to the scalar field, thus it
is local, justifying the use of the RG framework) have an
$N$-independent critical universal behavior.  This scenario is
supported by numerical results for $N=1,\,2,\,4,\,10,\,25$.  The MH
critical behavior does not apparently have an interpretation in terms
of the RG flow of the AHFT, as determined perturbatively close to four
dimensions or with standard large-$N$ methods.
\end{abstract}

\maketitle


\section{Introduction}
\label{intro}

Many emergent collective phenomena in condensed-matter
physics~\cite{Anderson-book,Wen-book} are explained by effective
three-dimensional (3D) scalar Abelian gauge models, in which scalar
fields are coupled with an Abelian gauge field. We mention the
transitions in superconductors~\cite{HLM-74,Herbut-book}, in quantum
$SU(N)$ antiferromagnets~\cite{RS-90, TIM-05, TIM-06, Kaul-12, KS-12,
  BMK-13, NCSOS-15, WNMXS-17,Sachdev-19}, and the unconventional
quantum transitions between the N\'eel and the valence-bond-solid
phases in two-dimensional antiferromagnetic $SU(2)$ quantum
systems~\cite{Sandvik-07, MK-08, JNCW-08, Sandvik-10, HSOMLWTK-13,
  CHDKPS-13, PDA-13, SGS-16}, which represent the paradigmatic models
for the so-called deconfined quantum criticality~\cite{SBSVF-04}. The
phase structure and the universal features of the transitions in
scalar gauge models have been extensively studied
~\cite{HLM-74,Herbut-book,FS-79,DH-81,FMS-81,DHMNP-81,CC-82,BF-83,
  FM-83,KK-85,KK-86,BN-86,BN-86-b,BN-87,RS-90,MS-90,KKS-94,BFLLW-96,
  HT-96,FH-96,IKK-96,KKLP-98,OT-98, CN-99, HS-00, KNS-02,MHS-02,
  SSSNH-02,SSNHS-03,MZ-03,NRR-03, MV-04,SBSVF-04, NSSS-04, SSS-04,
  HW-05, WBJSS-05, CFIS-05, TIM-05, TIM-06, CIS-06, KPST-06,
  Sandvik-07, WBJS-08, MK-08, JNCW-08, MV-08, KMPST-08, CAP-08, KS-08,
  ODHIM-09, LSK-09, CGTAB-09, CA-10, BDA-10, Sandvik-10, Kaul-12,
  KS-12, BMK-13, HBBS-13, Bartosch-13, HSOMLWTK-13, CHDKPS-13, PDA-13,
  BS-13, NCSOS-15, NSCOS-15, SP-15, SGS-16,WNMXS-17, FH-17, PV-19-CP,
  IZMHS-19, PV-19-AH3d, SN-19, Sachdev-19, PV-20-largeNCP, SZ-20,
  PV-20-mfcp, BPV-20-hcAH, BPV-21, BPV-21-bgs, BPV-21-ncAH, WB-21,
  BPV-22-mpf, BPV-22, BPV-23b,SZJSM-23, BPV-24}, paying particular
attention to the role of the gauge fields and of the related
topological features, like monopoles and Berry phases, which cannot be
captured by effective Landau-Ginzburg-Wilson (LGW) theories with
gauge-invariant scalar order
parameters~\cite{ZJ-book,PV-02,Sachdev-book,SBSVF-04,Sachdev-19}.

\begin{figure}[tbp]
\includegraphics*[width=0.95\columnwidth]{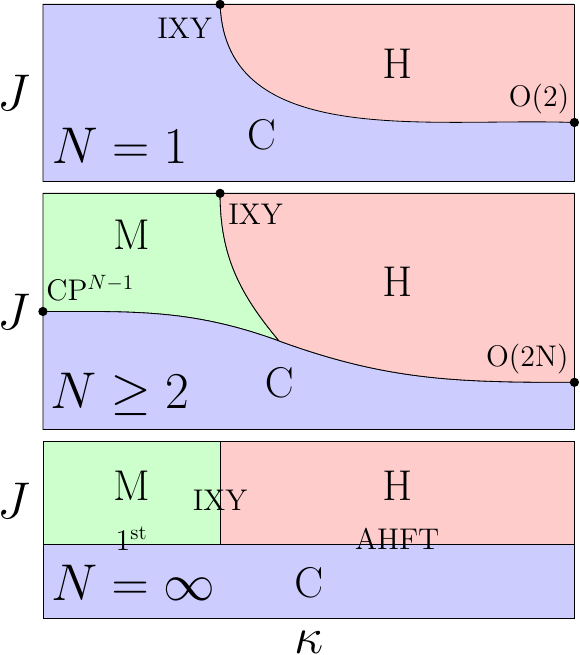}
  \caption{The $\kappa$-$J$ phase diagram of the $N$-component LAH
    model (\ref{AHH}), for $N=1$ (top), generic $N\ge 2$ (middle) and
    $N=\infty$ (bottom). For $N=1$, there are two phases, the Coulomb
    (C) and Higgs (H) phases, characterized by the confinement and
    deconfinement of charged gauge-invariant excitations,
    respectively.  For $N\ge 2$, the scalar field is disordered and
    gauge correlations are long ranged in the small-$J$ Coulomb (C)
    phase.  For large $J$ two phases occur, the molecular (M) and
    Higgs (H) ordered phase, in which the global $SU(N)$ symmetry is
    spontaneously broken.  The two phases are distinguished by the
    behavior of the gauge modes: the gauge field is long ranged in the
    M phase (small $\kappa$), while it is gapped in the H phase (large
    $\kappa$).  Moreover, while the C and M phases are confined
    phases, the H phase shows the deconfinement of charged
    gauge-invariant excitations.  }
\label{phdiasketchncLAH}
\end{figure}

Several lattice scalar gauge models have been considered, using both
compact and noncompact gauge variables, with the purpose of
identifying the possible universality classes of the continuous
transitions that occur in generic scalar gauge systems.  They provide
examples of topological transitions, which are driven by extended
charged excitations with no local order parameter, or by a nontrivial
interplay between long-range scalar fluctuations and nonlocal
topological gauge modes.

In this paper we address the topological deconfinement transitions
that occur in multicomponent lattice Abelian Higgs (LAH) models with a
global $SU(N)$ symmetry, in which a noncompact Abelian gauge field is
coupled with an $N$-component scalar field.

The phase diagrams of the 3D LAH models in the Hamiltonian parameter
space $J$-$\kappa$, see Eq.~(\ref{AHH}), are sketched in
Fig.~\ref{phdiasketchncLAH}. The phase diagram of the multicomponent
LAH models presents three phases, which differ in the properties of
the gauge correlations, in the confinement or deconfinement of the
charged excitations, and in the behavior under $SU(N)$
transformations.  In the small-$J$ Coulomb (C) phase, the scalar field
is disordered and gauge correlations are long ranged.  For large $J$
two phases occur, the molecular (M) and Higgs (H) ordered phase, in
which the global $SU(N)$ symmetry is spontaneously broken, the order
parameter being a gauge-invariant bilinear of the scalar field.  The
two phases are distinguished by the behavior of the gauge modes: the
gauge field is long ranged in the M phase (small $\kappa$), while it
is gapped in the H phase (large $\kappa$).  Moreover, while the C and
M phases are confined phases, the H phase shows the deconfinement of
charged gauge-invariant excitations, represented by nonlocal dressed
scalar operators \cite{Dirac:1955uv,KK-85,KK-86,BPV-23b}, whose
correlations do not vanish in the large-distance
limit~\cite{BPV-23b,KK-85,KK-86,BN-87}. For $N=1$, see the upper panel
of Fig.~\ref{phdiasketchncLAH}, there are only two phases, that
differ for the gauge behavior, as no global symmetry is present. As
it occurs for $N\ge 2$, charged scalar modes are deconfined in the H
phase~\cite{KK-85,KK-86,BN-87}.

In this paper we show that multicomponent LAH models can undergo
different types of {\em charged} deconfinement transitions, the CH
transition line between the C and H phases and the MH transition line
between the M and H phases. They are controlled by different {\em
  charged} fixed points of the renormalization-group (RG) flow with
nonzero gauge coupling.  Gauge correlations play an active, but
different, role at these deconfinement transitions and, therefore,
they cannot be described by effective LGW theories.

We focus on the MH transitions of multicomponent LAH models, which
have not been thoroughly analyzed yet.  While transitions along the CM
and CH lines are related to the spontaneous breaking of the global
$SU(N)$ symmetry, the MH line separates two ordered phases, both
characterized by the condensation of a gauge-invariant scalar-field
bilinear operator.  Therefore, the MH transitions must be driven by
the qualitative change of the gauge correlations, without a local
gauge-invariant order parameter, as it also occurs for the CH
transitions in the one-component LAH model, see
Fig.~\ref{phdiasketchncLAH}.

To investigate the critical behavior of the multicomponent LAH models
along the MH transition line and of the one-component LAH model along
the CH transition line, we report finite-size scaling (FSS) analyses
of Monte Carlo (MC) data for $N=1,\,2,\,4,\,10,\,25$. The results show
that these charged topological transitions are continuous, and their
critical behaviors belong to the same universality class, i.e., the
continuous MH transitions of the multicomponent LAH systems share the
same universality class of the CH transitions in the one-component LAH
model, see the upper panel of Fig.~\ref{phdiasketchncLAH}. Thus, gauge
correlations behave as in the so-called inverted $XY$ ($IXY$)
model~\cite{DH-81}, related to the standard $XY$ model by
duality~\cite{NRR-03}. Therefore, the critical behaviors of the
multicomponent LAH systems along the MH transition line differ from
those along the CH transition line, see Fig.~\ref{phdiasketchncLAH},
which are controlled by the stable fixed point of the 3D AH field
theory~\cite{BPV-21-ncAH,BPV-22,BPV-23b}.

The paper is organized as follows.  In Sec.~\ref{model} we present the
3D LAH model with noncompact gauge variables, we specify the
appropriate boundary conditions that ensure the absence of unphysical
divergences due to the gauge invariance of the model, and we define
the Lorenz gauge fixing we use to compute nongauge invariant gauge and
scalar correlations.  In Sec.~\ref{phasediag} we summarize the general
features of the phase diagram and of the transition lines.  In
Sec.~\ref{obsfss} we define the observables that we use in our
numerical analyses, and we report their expected FSS behavior.
Sec.~\ref{numres} reports the numerical results, i.e., the FSS
analyses of local and nonlocal gauge-invariant observables. Finally,
in Sec.~\ref{conclu} we summarize and draw our conclusions.

\section{The noncompact lattice Abelian Higgs theory}
\label{model}

We consider a LAH model with $N$-component complex vectors ${\bm
  z}_{\bm x}$ of unit length ($\bar{\bm z}_{\bm x} \cdot {\bm z}_{\bm
  x} =1$), and noncompact gauge variables $A_{{\bm x},\mu}\in {\mathbb
  R}$ ($\mu=1,2,3$).  The Hamiltonian and partition function are
(see, e.g., Refs.~\cite{KK-85,BN-86,BN-87,MV-04,BPV-21-ncAH})
\begin{eqnarray}
&&H = \frac{\kappa}{2} \sum_{{\bm x},\mu>\nu} F_{{\bm x},\mu\nu}^2 - 2NJ
  \sum_{{\bm x},\mu} {\rm Re}\,( \lambda_{{\bm x},\mu} 
  \bar{\bm z}_{\bm x} \cdot {\bm z}_{{\bm x}+\hat\mu}),\qquad
    \label{AHH}\\
  && Z = 
  \int [dA_{{\bm x},\mu} d\bar{\bm z}_{\bm z} d{\bm z}_{\bm x}]
  e^{-H({\bm A},{\bm z})},
\label{Zpart}
\end{eqnarray}
where $\lambda_{{\bm x},\mu} \equiv e^{iA_{{\bm x},\mu}}$, $F_{{\bm
    x},\mu\nu}\equiv\Delta_{\mu} A_{{\bm x},\nu} - \Delta_{\nu}
A_{{\bm x},\mu}$ and $\Delta_\mu A_{{\bm x},\nu} = A_{{\bm
    x}+\hat{\mu},\nu}- A_{{\bm x},\nu}$.  We have rescaled the
scalar-field coupling $J$ by a factor of $N$, to ensure that the limit
$N\to\infty$ at fixed $J$ is finite, see, e.g., Ref.~\cite{MZ-03}.
The
model has a global $SU(N)$ symmetry, ${\bm z}_{\bm x} \to V {\bm
  z}_{\bm x}$ with $V\in\mathrm{SU}(N)$, and a local Abelian gauge
invariance, 
\begin{eqnarray}
  {\bm z}_{\bm x} \to e^{i\Lambda_{\bm x}} {\bm z}_{\bm x},
  \qquad A_{{\bm x},\mu} \to A_{{\bm x},\mu} + \Lambda_{\bm
    x}-\Lambda_{{\bm x}+\hat{\mu}},
  \label{gaugeinv}
  \end{eqnarray}
  with $\Lambda_{\bm x}\in
\mathbb{R}$.

At variance with what happens for compact models, the partition
function (\ref{Zpart}) diverges, even on a finite lattice. This is due
to the existence of zero modes related with the gauge invariance
of the model.  This problem is not completely solved even by the use
of a maximal gauge fixing, if periodic boundary conditions are
chosen. With periodic boundary conditions the Hamiltonian $H$ is
indeed invariant under the group of noncompact transformations
$A_{{\bm x},\mu}\to A_{{\bm x},\mu} + 2\pi n_{\mu}$, where
$n_{\mu}\in\mathbb{Z}$ depends on the direction $\mu$ but is
independent of the point ${\bm x}$. This invariance is also (at least
partially) present in the gauge-fixed theory, and therefore $Z$ is
ill-defined also in this case.  To obtain a well-defined finite-volume
theory, we adopt $C^*$ boundary conditions~\cite{KW-91,
  LPRT-16,BPV-21-ncAH}. On a cubic lattice of size $L$, $C^*$ boundary
conditions are defined by
\begin{equation}
A_{{\bm x} + L\hat{\nu},\mu}= -A_{{\bm x},\mu},\quad
{\bm z}_{{\bm x}+L\hat{\nu}}= \bar{\bm z}_{{\bm x}}.
\label{Cstarbc}
\end{equation}
They preserve the local gauge invariance and softly break the $SU(N)$
global symmetry to O($N$), without affecting the bulk critical
behavior.

To compute some gauge and scalar correlation functions, we will
consider the Lorenz gauge fixing, defined by requiring
\begin{equation}
\sum_{\mu}\Delta_\mu^{-} A_{{\bm x},\mu}=0
\label{Lgauge} 
\end{equation}
for all lattice sites $\bm x$, where $\Delta_{\mu}^{-} A_{{\bm x},\nu}= A_{{\bm
x},\nu}-A_{{\bm x}-\hat{\mu},\nu}$. It breaks the invariance of the model
under the gauge transformations 
\begin{eqnarray}
A_{{\bm x},\mu}  &\to&  A'_{{\bm x},\mu} = A_{{\bm x},\mu} + \Lambda_{\bm x} -
    \Lambda_{{\bm x}+\hat{\mu}}, 
\nonumber \\
{\bm z}_{\bm x} &\to& {\bm z}'_{\bm x} = e^{i\Lambda_{\bm x}}{\bm z}_{\bm x}, 
\label{gaugetransf} 
\end{eqnarray}
where $\Lambda_{\bm x}$ is an arbitrary function of the lattice sites, that 
satisfies antiperiodic boundary conditions when $C^*$ boundary conditions
are adopted. 

As demonstrated in Ref.~\cite{BPV-23}, the lattice Lorenz gauge is
particularly convenient, as only zero-mode singularities occur in the
infinite-volume limit, at variance with what happens when working in
other lattice gauges, such as the axial gauge or the soft Lorenz
gauge. See Ref.~\cite{BPV-23b} for analogous considerations for the
scalar correlator.

Note that, for $N=1$, one could also use the so-called unitary gauge
that fixes $z_{\bm x} = 1$. The Hamiltonian becomes
\begin{equation}
      H_{\rm ug}= - 2 J 
\sum_{{\bm x},\mu} \cos A_{{\bm x},\mu}
+\frac{\kappa}{2} \sum_{{\bm x},\mu>\nu} F_{{\bm x},\mu\nu}^2.
\label{Hug}
\end{equation}
The unitary gauge fixing is not complete and indeed the Hamiltonian is
still invariant under gauge transformations in which $\Lambda_{\bm x}$
is a multiple of $2\pi$.  The model (\ref{Hug}) represents a soft
version of the $IXY$ gauge model that is obtained in the limit $J\to
\infty$ of the LAH models, see below.

\section{The phase diagram}
\label{phasediag}

In this section we summarize the general features of the $\kappa$-$J$
phase diagram of the 3D LAH models defined by the Hamiltonian
(\ref{AHH}). They show different features for $N=1$ and $N\ge 2$,
due to the possibility of the spontaneous breaking of the
global $SU(N)$ symmetry for $N\ge 2$, see Fig.~\ref{phdiasketchncLAH}.

\subsection{The one-component phase diagram}

For $N=1$ only two phases are present: a Coulomb (C) phase, in which
gauge correlators are gapless, and a Higgs (H) phase in which gauge
correlators are gapped; see, e.g., Ref.~\cite{BN-87}.  The C and
H phases can also be characterized by the confinement/deconfinement of
charged gauge-invariant excitations, represented by nonlocal dressed
scalar operators \cite{Dirac:1955uv,KK-85,KK-86,BPV-23b}, whose
correlation functions do not vanish in the large-distance
limit~\cite{BPV-23b,KK-85,KK-86,BN-87} in the H phase.

The C and H phases are separated by a transition line connecting the
transition points occurring in the $J\to\infty$ and $\kappa\to\infty$
limits, where the noncompact LAH model becomes equivalent to the $IXY$
model and to the standard O(2)-vector spin model, respectively.  For
$J\to\infty$ the gauge field $A_{{\bm x},\mu}$ takes only values which
are multiples of $2\pi$. Indeed, the $J\to\infty$ limit leads to the
constraints
\begin{eqnarray}
{\bm{z}}_{\bm x} =
\lambda_{{\bm x},\mu}\, {\bm z}_{{\bm x}+\hat\mu},\quad \lambda_{{\bm
    x},{\mu}} \,\lambda_{{\bm x}+\hat{\mu},{\nu}}
\,\bar{\lambda}_{{\bm x}+\hat{\nu},{\mu}} \,\bar{\lambda}_{{\bm
    x},{\nu}} = 1.
\label{constraints}
\end{eqnarray}
Then, by an appropriate gauge transformation, one can set $A_{{\bm x},
  \mu} = 2 \pi n_{{\bm x}, \mu}$, where $n_{{\bm x}, \mu} \in {\mathbb
  Z}$. The resulting $IXY$ gauge model is a dual-loop representation
of the 3D $XY$ model, more precisely its free energy is related by
duality to that of the $XY$ model with Villain
action~\cite{DH-81,NRR-03}.  This dual-loop model undergoes an $IXY$
transition, i.e., a transition belonging to the $XY$ universality
class, but with inverted high- and low-temperature
phases~\cite{DH-81}.  Moreover, in the dual-loop model only thermal RG
operators perturb the $XY$ fixed point (no magnetic perturbations
breaking the global $U(1)$ symmetry are present).  Therefore, the 3D
LAH models must undergo a $J\to\infty$ $IXY$ transition point located
at $\kappa_c(J\to\infty) = 0.076051(2)$~\cite{NRR-03,BPV-21-ncAH}.

In the limit $\kappa\to\infty$, all plaquettes $F_{{\bm x},\mu\nu}$
vanish and thus the model is equivalent (up to a gauge transformation)
to the $XY$ model. Therefore, $\kappa\to\infty$ the one-component LAH
model is expected to undergo an $XY$ transition
at~\cite{Hasenbusch-19,CHPV-06,DBN-05}
$J_c(\kappa\to\infty)=0.22708234(9)$ (this $XY$ transition is denoted
by O(2) in Fig.~\ref{phdiasketchncLAH}).

\subsection{The multi-component phase diagram}
\label{multicompha}

The $\kappa$-$J$ phase diagram of the multicomponent LAH models (see
Refs.~\cite{MV-08,KMPST-08,BPV-21-ncAH}) is sketched in the middle
panel of Fig.~\ref{phdiasketchncLAH}. For $N\ge 2$ the model is also
invariant under $SU(N)$ global transformations. Thus, transitions
associated with the breaking of the $SU(N)$ symmetry and phases
characterized by standard, i.e., nontopological, order can also be
present. The breaking of the $SU(N)$ symmetry can be characterized by
using the gauge-invariant order parameter
\begin{equation}
  Q_{{\bm x}}^{ab} = \bar{z}_{\bm x}^a z_{\bm x}^b -\delta^{ab}/N,
  \label{Qdef}
\end{equation}
which transforms in the adjoint representation of the $SU(N)$ global
symmetry group.

The phase diagram for $N\ge 2$ is characterized by three different
phases.  For small $J$ there is a Coulomb phase (C), which is $SU(N)$
symmetric ($Q^{ab}_{\bm x}$ is disordered) and in which the gauge
field is gapless. For large $J$ values there are two phases in which
the $SU(N)$ symmetry is broken ($Q^{ab}_{\bm x}$ condenses).  They are
characterized by the different behavior of the gauge modes: in the
molecular phase (M) the gauge field is long ranged (as in the C
phase), while in the Higgs phase (H) it is short ranged.  From the
point of view of the gauge-invariant charged excitations, the C and M
phases are confined, while the H phase is deconfined.

The existence of these three phases is consistent with the analysis of
the model behavior for large and/or vanishing values of $J$ and
$\kappa$.

(i) For $J=0$ the LAH model reduces to the three-dimensional
$CP^{N-1}$ model, which is known to undergo a continuous $O(3)$
transition for $N=2$ and discontinuous transitions for $N>3$
\cite{PV-19-CP}. Therefore, along the $\kappa=0$ line, we have a
continuous O(3) transition point located at $J_c(\kappa=0)= 0.7102(1)$
for $N=2$, and first-order transitions for $N>2$ (for example at
$J_c(\kappa=0) \approx 0.353$ for $N=20$), see e.g.
Refs.~\cite{PV-19-CP,PV-20-largeNCP}.

(ii) For $J\to\infty$ the multicomponent model behaves as 
the one-component model~\cite{BPV-21-ncAH}.  Indeed, 
Eq.~(\ref{constraints}) holds for any $N$, and therefore in all cases
the gauge field $A_{{\bm x},\mu}$ takes only values
which are multiples of $2\pi$ and the scalar field decouples.
Therefore, we have a $J\to\infty$ $IXY$ transition 
for $\kappa=\kappa_c(J\to\infty) =
0.076051(2)$~\cite{NRR-03,BPV-21-ncAH}, independently of $N$.

(iii) For $\kappa=\infty$ all plaquettes vanish and the model reduces,
up to a gauge transformation, to the standard 
O(2$N$) vector model. For example, we must have $J_c(\kappa\to\infty)
= 0.23396363(6)$ for $N=2$~\cite{Hasenbusch-22}, and
$J_c(\kappa\to\infty) = J_{c,\infty} + a_{1} N^{-1} + O(N^{-2})$ in
the large-$N$ limit, with $J_{c,\infty}=0.252731...$ and $a_1 \approx
-0.234$~\cite{CPRV-96}.

\subsection{Critical behaviors along the transition lines}
\label{translines}

The three phases of the multicomponent LAH model are separated by
three different transition lines, the CM, CH, and MH transition lines,
which start from the transition points located at $\kappa=0$,
$J=\infty$ and $\kappa=\infty$. The transitions
along the lines separating the different phases may be of first order
or continuous and, in the latter case, belong to universality classes
that may depend on the number $N$ of scalar components.  The
continuous transitions are related to the stable (charged or
uncharged) fixed points of the RG flow, each one with its own
attraction domain in the model parameter space.

The CM and CH transition lines of the multicomponent LAH models have
already been thoroughly investigated. The CM transitions are in the
same universality class as that of the 3D $CP^{N-1}$ model (defined on
the line $\kappa=0$). An effective description is provided by a LGW
model without gauge
fields~\cite{BPV-21-ncAH,PV-19-CP,PV-19-AH3d,PV-20-largeNCP}.  The
stable fixed point is {\em uncharged}, and gauge fields have only the
role of hindering non-gauge-invariant modes from becoming critical.
For $N\ge 3$ the LGW theory predicts a generic first-order transition,
while, for $N=2$, the transitions can be continuous in the O(3) vector
universality class.  

The continuous CH transitions are associated with
the stable charged fixed point of the RG flow of the AH field theory
(AHFT)~\cite{HLM-74,IKK-96,MZ-03,KS-08,IZMHS-19,BPV-21-ncAH,BPV-22,BPV-23b}
\begin{equation}
  {\cal L} = \frac{1}{4 g^2} \,F_{\mu\nu}^2 +
|D_\mu{\bm\Phi}|^2 + r\, {\bm \Phi}^*{\bm \Phi} + \frac{1}{6} u
\,({\bm \Phi}^*{\bm \Phi})^2 
\label{AHFT}
\end{equation}
($F_{\mu\nu}\equiv \partial_\mu A_\nu - \partial_\nu A_\mu$ and $D_\mu
\equiv \partial_\mu + i A_\mu$), which corresponds to the formal
continuum limit of the LAH model (\ref{AHH}), relaxing the unit-length
constraint for the scalar field. Continuous charged CH transitions in
the 3D LAH model occur for $N>N^\star$ with
$N^\star=7(2)$~\cite{BPV-21-ncAH,BPV-22}. 
Critical
exponents depend on $N$, consistently with the large-$N$ field-theory
predictions~\cite{HLM-74,IKK-96,MZ-03,KS-08,BPV-21-ncAH,BPV-23b}.  The
O($2N$) vector fixed point for $g=0$ (corresponding to
$\kappa\to\infty$ in the LAH model) is unstable with respect to 
gauge fluctuations for any $N$~\cite{BPV-21-ncAH}.

The transitions along the MH line have not been thoroughly analyzed
yet.  The MH line separates two ordered phases, both characterized by
the condensation of the gauge-invariant bilinear $Q_{\bm x}^{ab}$
defined in Eq.~(\ref{Qdef}). Therefore, the MH transitions must be
related to the qualitative change of the gauge correlations, without a
local gauge-invariant order parameter, as it also occurs for the CH
transitions in the one-component LAH model.  In the following we show
that these topological transitions are continuous, at least for
sufficiently large (but finite) values of $J$, and controlled by
another charged fixed point, different from the AHFT fixed point that
controls the CH transitions.  As we shall see, the MH fixed point
turns out to be the same as that controlling the $J\to\infty$ $IXY$
transition and the continuous CH transitions in the $N=1$ LAH model,
see the upper panel of Fig.~\ref{phdiasketchncLAH}.

The existence of the MH transition line for $N\ge 2$, see
Fig.~\ref{phdiasketchncLAH}, can be inferred from the presence of two
low-temperature phases distinguished by the nature of the gauge
correlations. This is already suggested by the $IXY$ transition in the
$J\to\infty$ limit. A natural hypothesis is that the $J\to\infty$
$IXY$ transition point is the starting point of the MH transition line
for $N\ge 2$ and of the CH line for $N=1$, see
Fig.~\ref{phdiasketchncLAH}.

We wish to understand whether the MH transitions belong to the same
universality class as the $IXY$ transition that occurs for
$J\to\infty$. This identification is not obvious. Indeed, for $N\ge
2$, in the M and H phases scalar fluctuations are only partially
frozen, because of the presence of $2N-2$ massless Goldstone bosons
related with the spontaneous breaking of the $SU(N)$ symmetry, from
$SU(N)$ to $U(N-1)$~\cite{MZ-03}.  Therefore, the multicomponent
scalar fluctuations may be relevant, giving rise to an $N$-dependent
critical behavior.  In this case, the fluctuations of the scalar field
would drive the system toward a different asymptotic behavior, giving
rise to first-order transitions or to a different critical behavior
associated with a more stable charged fixed point. Nonetheless, it is
also possible that the residual scalar fluctuations, and in particular
the long-range Goldstone modes, are irrelevant at the MH transitions,
somehow decoupling from the topological gauge-field critical modes
(this scenario was originally mentioned in Ref.~\cite{MV-08} as a
plausible hypothesis, without providing evidence). In this case, the
critical behavior along the MH line would be the same as that along
the CH line for $N=1$, where scalar fields can be eliminated by a
gauge transformation (unitary gauge) and therefore $IXY$ critical
behavior for finite $J$ arises naturally.

\subsection{The large-$N$ phase diagram}
\label{largeN}

For $N=\infty$ the geometry of the phase diagram is simpler.  First,
we argue that the MH line is a straight line corresponding to $\kappa
= \kappa_{c,IXY}$. Indeed, since $N$ and $J$ appear in the combination
$NJ$, the $N\to\infty$ limit is somewhat similar to the $J\to\infty$
limit. However, they are not equivalent, since, by changing $N$, one
also changes the number of components of the scalar field.  We shall
now argue the this equivalence holds in the M and H phases. Indeed, in
this case the $SU(N)$ symmetry is broken. We consider magnetized
boundary conditions, i.e., we set $z_{\bm x} = e^{i\alpha_x} {\bm
  e}_1$, ${\bm e}_1= (1,0,\ldots)$ on the boundary, where
$e^{i\alpha_x}$ is an unconstrained phase that guarantees that the
boundary conditions do not break the gauge invariance of the
model. Since the $SU(N)$ symmetry is broken in the M and H phases, in
the bulk we expect $z_{\bm x} = z_\parallel {\bm e}_1 + {\bm
  z}_\perp$, with $|z_\parallel|$ approximately equal on all lattice
sites.  As the number of components $N$ increases, we expect ${\bm
  z}_{{\bm x} \perp}\cdot {\bm z}_{{\bm y} \perp}$ on neighbouring
sites to decrease to zero, so that the scalar Hamiltonian would
converge to
\begin{equation}
- 2 N J \sum_{{\bm x} \mu} \hbox{Re}\, \bar{z}_{{\bm x}, \parallel} 
{z}_{{\bm x} + \hat{\mu}, \parallel} \lambda_{x,\mu}, 
\end{equation}
in terms of the single-component quantity $z_\parallel$. It follows 
that the two limits ($N\to \infty$ and $J\to\infty$) should be
equivalent, implying the independence of the MH line on $J$. The
behavior along the CH line can be obtained by using the AHFT, since
these transitions are controlled by the field-theory fixed point.  The AHFT 
predicts the large-$N$ behavior to be independent of the gauge 
fields~\cite{MZ-03}, and the same should hold for the lattice model.
Therefore, we predict the CH line to
be a straight line with $J = J_{c,\infty} = 0.252731...$, where
$J_{c,\infty}$ is the values of $J$ where the O($2N$) transition
($\kappa = \infty$) occurs.  The shape of the CM line is less clear,
given that standard large-$N$ lattice calculations are not reliable
for the LAH model in the $\kappa\to 0$ limit (i.e., for the $CP^{N-1}$
model) \cite{PV-20-largeNCP}.  Numerical simulations indicate that the
large-$N$ transitions are of first order for any $N\ge 3$, and that
they become stronger and stronger with increasing
$N$~\cite{BPV-21-ncAH,PV-20-largeNCP}. If we accept the conjecture
(supported by numerical data) of Ref.~\cite{PV-20-largeNCP} that the
$CP^{N-1}$ transition ($\kappa =0$) occurs at the same value
$J_{c,\infty} = 0.252731...$ where the O($2N$) transition ($\kappa =
\infty$) occurs, we can conjecture that also the CM transition line
corresponds to the line $J=J_{c,\infty}$ for all values of
$\kappa$. We thus obtain the simple phase diagram shown in the bottom
Fig.~\ref{phdiasketchncLAH}.  In this case the multicritical point
would be located at $J=J_{c,\infty}$ and $\kappa=\kappa_{c,IXY}$.

\section{Observables and Finite-size scaling}
\label{obsfss}

\subsection{Observables}
\label{defobs}

Most investigations of the multicomponent LAH model studied the
critical behavior of correlations of the gauge-invariant bilinear operator
$Q_{\bm x}^{ab}$, which characterizes the $SU(N)$ symmetry breaking
and is therefore an appropriate order parameter for the CM and CH transitions
(see, e.g., Ref.~\cite{BPV-21-ncAH, BPV-23b}). This gauge-invariant
observable is not relevant for the MH transitions, as
the $SU(N)$ symmetry is broken in both phases. We need therefore a
different set of observables to characterize the critical behavior.

\subsubsection{The gauge-invariant energy cumulants}
\label{ecum}

In our FSS analyses we consider the gauge-invariant energy cumulants
$B_k$, which are intensive quantities related to the energy central
moments
\begin{equation}
  M_k = \langle \, (H-\langle H\rangle)^k \, \rangle,
  \label{mkmom}
\end{equation}  
by
\begin{eqnarray}
  & B_1=L^{-3}\langle H\rangle,\quad   & B_2=L^{-3} M_2, \nonumber \\
  & B_3 = L^{-3} M_3, \quad   & B_4 = L^{-3} (M_4 - 3 M_2^2),\qquad
\end{eqnarray}
etc. Note that $B_2$ is proportional to the specific heat.  These
global quantities allow one to characterize topological transitions in
which no local gauge-invariant order parameter is present, see, e.g.,
Refs.~\cite{SSNHS-03,BPV-20-hcAH,BPV-22-z2g}.

\subsubsection{Gauge-field correlations in the Lorenz gauge}

To determine gauge-field correlations, we first define gauge-dependent
correlation functions in the Lorenz gauge. In the next Section
we will show that these quantities provide information on the critical
behavior on a set of gauge-invariant correlators. 
We start by defining the Fourier-transformed field
$\widetilde{A}_\mu({\bm p})$,
\begin{equation}
\widetilde{A}_\mu({\bm p}) = e^{i p_\mu/2} \sum_{\bm x} e^{i{\bm p}\cdot{\bm x}}
  A_{{\bm x},\mu},
\label{tildeA}
\end{equation}
where the prefactor takes into account that the gauge field is
naturally defined on the lattice links and guarantees that
$\widetilde{A}_\mu({\bm p})$ is odd under reflections in momentum
space.  The correlation function is defined as
\begin{eqnarray}
  \widetilde{C}_{\mu\nu}({\bm p}) = L^{-3} \, \langle
  \widetilde{A}_\mu({\bm p}) \widetilde{A}_\nu(-{\bm p})\rangle.
\label{twopA}
\end{eqnarray}
The momenta ${\bm p}$ run over the values $p_i = \pi (2 n_i + 1)/L$
with $ n_i = 0,\ldots L-1$, since $A_{{\bm x},\mu}$ is antiperiodic
due to the $C^*$ boundary conditions. In particular, ${\bm p} = 0$ is
not allowed.  Note that $\langle A_{{\bm x},\mu}\rangle=0$, since the
charge-conjugation symmetry $A_{{\bm x},\mu}\to -A_{{\bm x},\mu}$ is
preserved both by the $C^*$ boundary conditions and by the Lorenz
gauge.

The gauge-field susceptibility is defined as
\begin{eqnarray}
  \chi_{A} = \widetilde{C}_{\mu\mu}({\bm p}_a),
   \label{chiAdef}
\end{eqnarray}
where $\mu$ is one of the directions (no sum on repeated indices
implied) and ${\bm p}_a$ is one of the smallest momenta compatible
with the antiperiodic boundary conditions:
\begin{equation}  \label{padef}
{\bm p}_a=(\pi/L, \pi/L, \pi/L).
\end{equation}
The second-moment correlation length of the gauge field is defined by
\begin{equation}
\label{xiA}
\xi^2_{A} = 
\frac{1}{(\hat{p}_a^2-\hat{p}_b^2) }
\frac{\widetilde{C}_{\mu\mu}({\bm p}_b)
  - \widetilde{C}_{\mu\mu}({\bm p}_a)}
{\widetilde{C}_{\mu\mu}({\bm p}_a)},
\end{equation}
where  
\begin{equation}
  \hat{p}^2=\sum_{\mu=1}^3 4\sin^2(p_\mu/2),
  \qquad
  {\bm p}_b={\bm p}_a+{2\pi \over L} \hat{\nu},
\end{equation}
and, somewhat arbitrarily, we have taken $\nu\not=\mu$.  Any pair of
directions $\mu,\nu$ are obviously equivalent.  The Binder cumulant of
the gauge field is instead defined by
\begin{eqnarray}
  U_{A} = \frac{\langle m_{2,\mu}^2\rangle}{\langle m_{2,\mu} \rangle^2},
  \qquad
m_{2,\mu} = 
\big|\sum_{\bm x} e^{i{\bm p}_a\cdot {\bm x}} A_{{\bm x},\mu}\big|^2.
    \label{binderdef2}
\end{eqnarray}

\subsubsection{Gauge-invariant correlators of the gauge field}
\label{gau}

Let us now show that the gauge-dependent quantities defined in the previous 
Section allow us to determine the critical behavior of gauge-invariant
plaquette correlations. 
Indeed, let us define the gauge-invariant correlator
\begin{equation}
C_{F,\mu\nu,\alpha\beta}({\bm p} )
 = {1\over V} \langle 
    \widetilde{F}_{\mu\nu}(-{\bm p})
   \widetilde{F}_{\alpha\beta} ({\bm p}) \rangle,
\end{equation}
where $\widetilde{F}_{\alpha\beta} ({\bm p})$ is the Fourier transform
of the plaquette operator $F_{{\bm x},\mu\nu}$:
\begin{equation}
\widetilde{F}_{\mu\nu}({\bm p}) = e^{i (p_\mu + p_\nu)/2} \sum_{\bm x} 
e^{i{\bm p}\cdot {\bm x}} F_{{\bm x},\mu\nu}.
\end{equation}
It is simple to relate correlations of this gauge-invariant operator
to correlations of $A_{{\bm x},\mu}$ computed in the Lorenz gauge. For
instance, we have
\begin{equation}
\sum_{\mu\nu} 
   C_{F,\mu\nu,\mu\nu} ({\bm p}) = 
2 \hat{p}^2 \sum_\nu \langle \widetilde{A}_{\nu}(-{\bm p})
   \widetilde{A}_{\nu} ({\bm p}) \rangle.
\end{equation}
From this relation it immediately follows that the susceptibility
$\chi_A$ of the field $A_{{\bm x},\mu}$ is proportional to
$L^2\chi_F$, where $\chi_F$ is the plaquette susceptibility; more
precisely, we have for large values of $L$
\begin{eqnarray}
\chi_A = {L^2\over 18 \pi^2} \chi_F, \qquad
\chi_F =  \sum_{\mu\nu}
   C_{F,\mu\nu,\mu\nu} ({\bm p}_a). \label{chifrel}
\end{eqnarray}
Also $\xi_A$ can be related to particular correlations of the
plaquette operator. In particular, choosing $\mu = 1$ and $\nu = 3$ in
the definition (\ref{xiA}), for large values of $L$ we have the
relations
\begin{eqnarray}
\widetilde{C}_{11}({\bm p}_a) &=& {L^2 \over 18 \pi^2} \chi_F, \\
\widetilde{C}_{11}({\bm p}_b) &=& {L^2 \over 121 \pi^2} 
    [10 C_{F,12,12}({\bm p}_b) + 9 C_{F,13,13}({\bm p}_b)]. 
\nonumber 
\end{eqnarray}

\subsubsection{Scalar-field correlations in the Lorenz gauge}

Scalar-field observables can be defined analogously, by setting
\begin{equation}
  \widetilde{\bm z}({\bm p}) = \sum_{\bm x}
  e^{i{\bm p}\cdot {\bm z}} {\bm z}_{\bm x}, \qquad 
\widetilde{G}_z({\bm p}) = {1\over L^3}
    \left\langle \left| \widetilde{\bm z}({\bm p}) \right|^2 \right\rangle.
\label{gxypz}
\end{equation}
The corresponding susceptibility $\chi_z$ and length scale $\xi_z$ are
defined as
\begin{eqnarray}
\chi_z = \widetilde{G}_z({\bm 0}),
\quad
\xi_z^2 \equiv {1\over 4 \sin^2 (\pi/L)} {\widetilde{G}_z({\bm 0}) -
   \widetilde{G}_z({\bm p}_m)\over \widetilde{G}_z({\bm p}_m)},\quad
\label{xidefpbz}
\end{eqnarray}
where ${\bm p}_m = (2\pi/L,0,0)$ has been selected quite arbitrarily,
since ${\bm z}_{\bm x}$ is neither periodic nor antiperiodic (see
Ref.~\cite{BPV-23b} for a more detailed discussion of this issue). The
Binder cumulant for the scalar field is defined by
\begin{eqnarray}\label{binderdefz}
  U_{z} = 
\frac{\langle m_{2}^2\rangle}{\langle m_{2} \rangle^2}, \qquad
m_{2} = \sum_{{\bm x},{\bm y}} \bar{\bm z}_{\bm x} \cdot {\bm z}_{\bm y}.
\end{eqnarray}

\subsubsection{Gauge-invariant correlations of
  nonlocal charged operators}
\label{equinonloc}

We now show that the above defined gauge-dependent scalar quantities
correspond to gauge-invariant observables computed in the Lorenz
gauge. To investigate the behavior of charged quantities, one can
consider the gauge-invariant operator ${\bm \Gamma}_{\bm x}$ defined
by~\cite{Dirac:1955uv}
\begin{equation}
\begin{aligned}
& {\bm \Gamma}_{\bm x} = {\bm z}_{\bm x}
  \exp\left(i \sum_{{\bm y},\mu} E_\mu({\bm y},{\bm x}) 
    A_{{\bm y},\mu}\right) , \label{phixdef}\\
& E_{\mu}({\bm y},{\bm x}) = V({\bm y}+\hat{\mu},{\bm x}) -
V({\bm y},{\bm x}).
\end{aligned}
\end{equation}
In this expression $V(\bm{x},{\bm y})$ is the lattice Coulomb
potential in ${\bm x}$ due to a unit charge in ${\bm y}$, i.e., the
solution of the lattice equation
\begin{equation}
  \sum_\mu \Delta^-_\mu \Delta_\mu V({\bm x},{\bm y}) = -
   \delta_{{\bm x},{\bm y}},
\end{equation}
in which the lattice derivatives act on the ${\bm x}$ variable.  It is
easy to verify that the lattice Poisson equation always has a unique
solution when $C^*$ boundary conditions are used, unlike the case of
periodic boundary conditions.  The operator ${\bm \Gamma}_{\bm x}$ is
invariant under the local gauge transformations
(\ref{gaugetransf}). It is enough to note that
$\Delta_{\mu}^{\dag}=-\Delta_{\mu}^-$, so that
\begin{eqnarray}
\sum_{{\bm y},\mu} E_\mu({\bm y},{\bm x}) \Delta_\mu \Lambda_{\bm y} &= &
-\sum_{{\bm y},\mu} \Delta_\mu^- E_\mu({\bm y},{\bm x}) \Lambda_{\bm y} = 
\nonumber \\
&=& \sum_{{\bm y}} \delta_{{\bm y},{\bm x}} \Lambda_{\bm y} = \Lambda_{\bm x}.
\end{eqnarray}
On the other hand, under the global U(1) transformation ${\bm z}_{\bm
  x} \to e^{i\alpha} {\bm z}_{\bm x}$ (which is not an allowed gauge
transformation when $C^*$ boundary conditions are used), the operator
${\bm \Gamma}_{\bm x}$ transforms as ${\bm \Gamma}_{\bm x} \to
e^{i\alpha} {\bm \Gamma}_{\bm x}$, and thus is a \emph{charged}
gauge-invariant operator. In the Lorenz gauge
\begin{equation}
\sum_{{\bm y},\mu} E_\mu({\bm y},{\bm x}) A_{{\bm y}, \mu}  = 
-\sum_{{\bm y},\mu} V({\bm y},{\bm x}) \Delta^{-}_{\mu} A_{{\bm y},\mu}=0, 
\end{equation}
so that ${\bm \Gamma}_{\bm x}$ reduces to ${\bm z}_{\bm
  x}$. Therefore, correlation functions of ${\bm \Gamma}_{\bm x}$, such as
\begin{equation}
G_\Gamma({\bm x},{\bm y}) = \langle \bar{\bm \Gamma}_{\bm x}
\cdot {\bm \Gamma}_{\bm y}\rangle
\label{GGamma}
\end{equation}
can be computed as correlation functions of ${\bm z}_{\bm x}$,
i.e. $G_z({\bm x},{\bm y}) = \langle \bar{\bm z}_{\bm x} \cdot
{\bm z}_{\bm y}\rangle$ in the Lorenz gauge.  We recall that, as
demonstrated for $N=1$~\cite{KK-85,KK-86,BN-86,BN-86-b} and
numerically confirmed for multicomponent systems~\cite{BPV-23b}, 
the charged excitations associated with ${\bm \Gamma}_{\bm
  x}$ condense in the H phase, i.e.,  $G_\Gamma({\bm x},{\bm y})\to c
\neq 0$ (equivalently $G_z({\bm x},{\bm y})\to c \neq 0$ in the Lorenz gauge)
in the large $|{\bm x}-{\bm y}|$ limit.

It is interesting to observe that the correlations of the operator
${\bm \Gamma}_{\bm x}$ converge to correlations of the gauge nonlocal
operator
\begin{equation}
 \widetilde{\Gamma}_{\bm x} = 
  \exp\left(i \sum_{{\bm y},\mu} E_\mu({\bm y},{\bm x}) 
    A_{{\bm y},\mu}\right) 
\label{phixdef-IXY}
\end{equation}
in the limit $J\to \infty$, i.e., in the $IXY$ model. The operator
$\widetilde{\Gamma}_{\bm x}$ is invariant (the exponents vary by
multiples of $2\pi i$) under the restricted gauge transformations that
are appropriate for the $IXY$ model. Note that this identification
allows us to conclude that the critical behavior of
$\widetilde{\Gamma}_{\bm x}$ in the $IXY$ model (no scalar fields are
present here) is the same as that of the scalar-field correlations in
the LAH model with Lorenz gauge fixing.

\subsection{Finite-size scaling}
\label{fssa}

We summarize here the main FSS relations that we exploit in our
numerical analysis. 
We consider simulations varying $\kappa$ at fixed $J$, so that 
the basic FSS variable is 
\begin{eqnarray}
  X = (\kappa - \kappa_{c}) L^{1/\nu},
  \label{Xdef}
\end{eqnarray}
where $\kappa_c$ is the critical value, $\nu$ is the
length-scale critical exponent, and $L$ is the lattice size.

RG invariant quantities, such as the ratios
\begin{equation} 
R_{A} = \xi_{A}/L, \qquad R_{z} = \xi_z/L, 
\end{equation}
and the Binder parameters $U_{A}$, $U_z$, scale in the large-$L$ limit
as
\begin{equation}
\label{eq:FSS1}
R(\kappa,L)={\cal R}(X)+ O(L^{-\omega}),
\end{equation}
where ${\cal R}$ is universal apart from a normalization of the
argument $X$ and $\omega$ is the leading correction-to-scaling
exponent. The scaling relation Eq.~\eqref{eq:FSS1} can be written in
a different (and often more useful) way when two RG-invariant
quantities $R$ and $R_1$ are available, and $R_1$ is monotonic with
respect to $\kappa$, as it occurs for the ratios $R_{A}$ and $R_{z}$.
In this case we can replace $X$ with $R_1$ and write the asymptotic
FSS behavior as
\begin{equation}
\label{eq:FSS2}
R(\kappa,L)=\widehat{\cal R}(R_1) + O(L^{-\omega}),
\end{equation}
where $\widehat{\cal R}$ is a universal function with no nonuniversal
normalization. The FSS relation (\ref{eq:FSS2}) is particularly useful
to check universality, since it does not require any parameter tuning.

To determine the scaling dimension of a local operator $O_{\bm x}$, we
analyze the corresponding susceptibility $\chi_O$, which can defined
in terms of the Fourier-transformed two-point correlation function at
small momentum, as in Eqs.~\eqref{chiAdef}, \eqref{xidefpbz}. In the
FSS limit, the susceptibility $\chi_O$ behaves as
\begin{equation}
  \chi_{O}\approx L^{d-2 y_O} F_{\chi}(R_1) = L^{2-\eta_O}
  F_{\chi}(R_1),
\label{chiosca}
\end{equation}
where $F_\chi$ is a function which is universal apart from a
multiplicative factor, and we used the standard RG relation
$y_{O}=(d-2+\eta_{O})/2$. Note that Eq.~\eqref{chiosca} gives the
leading large-size behavior for $\kappa\approx \kappa_c$ only if
$y_{O} < d/2$ (or, equivalently, if $\eta_O<2$). If this is not the
case, the analytic background is the dominant contribution and the
nonanalytic scaling part represents a correction term.

The cumulants $B_k$ are expected to show the FSS
behavior~\cite{SSNHS-03,BPV-20-hcAH,BPV-22-z2g}
\begin{equation} 
B_k(\kappa,L) \approx L^{k/\nu-3} \left[ {\cal B}_k(X) +
  O(L^{-\omega})\right] + b_k,
\label{Hg3-scaling}
\end{equation}
where the constant $b_k$ represents the analytic
background~\cite{PV-02,BPV-20-hcAH}. The scaling functions ${\cal
  B}_k(X)$ are universal apart from a multiplicative factor and a
normalization of the argument. We recall that they generally depend on
the chosen boundary conditions.

Some important remarks are in order before applying the FSS approach
to determine the universal features of the MH transitions.  If the
critical behavior is the same as in the $IXY$ model, which, in turn,
is related by duality to the standard $XY$ model, we should have
~\cite{CHPV-06,Hasenbusch-19,CLLPSSV-20,GZ-98,PV-02,KP-17},
$\nu=\nu_{XY}=0.6717(1)$ and $\omega=\omega_{XY} \approx 0.79$. In
this case, $B_2$ is not convenient, as the leading behavior is
dominated by the constant $b_2$, due to the fact that
$\alpha=2-3\nu<0$. Thus, we focus on the third cumulant that diverges
with exponent $3/\nu-3\approx 1.47$.

It is important to stress that the Lorenz-gauge representation of the
charged operator $\Gamma_{\bm x}$ in terms of the {\em local} scalar
field ${\bm z}_{\bm x}$ allows us to use the standard RG
framework~\cite{Wilson-83,Fisher-75,Wegner-76,Privman-90,PV-02}, to
predict the critical behavior of its correlations. This allows us to
write down power-law FSS behaviors analogous to those valid for
correlations of local operators. As we shall see, this allows us to
characterize their critical behavior, showing that their power laws
are controlled by a new universal critical exponent, which turns out
not to depend on the number of components.

Finally, we note that the relation (\ref{chifrel}), which is valid for
any $J$ and, in particular, in the limit $J\to\infty$, allows us to
prove $\eta_A = 1$ in the $IXY$ model.  Ref.~\cite{NRR-03} explicitly
showed that $\chi_F$ in the $IXY$ model is related by a duality
transformation to the helicity modulus $\Upsilon$ in the Villain $XY$
model, i.e., $\chi_F \sim \Upsilon$. Since $ \Upsilon \sim L^{-1}$ at
an $XY$ critical point, see Ref.~\cite{FBJ-73}, we obtain
\begin{equation}
\chi_A\sim L^2\chi_F\sim L^2 \Upsilon \sim L.
\end{equation}
The exponent $\eta_A$ can be determined from the large-size behavior
of $\chi_A$, as $\chi_A\sim L^{2-\eta_A}$.  We thus conclude that
$\eta_A=1$ in the $IXY$ model. We will show below that this result
extends to all MH transitions for any value of $N$.

\section{Numerical results}
\label{numres}

\subsection{Monte Carlo simulations}
\label{MCsim}

To investigate the nature of the critical behavior along the MH
transition line, and to compare it with the behavior observed along
the $N=1$ CH line, we present numerical FSS analyses of data obtained
by MC simulations, considering cubic lattices of size $L^3$ with $C^*$
boundary conditions, defined in Eq.~(\ref{Cstarbc}).

We have performed MC simulations at fixed $J$, varying $\kappa$ close
to the MH (CH for $N=1$) transition line. We have obtained results for
$N=1, 2, 4, 10, 25$ along the line $J=1$, considering lattices of size
$L$ up to 26, 26, 20, 20, 20 respectively. For $N=25$, we have also
performed simulations along the line $J=0.4$, with $L$ up to 32.  For
comparison note the transition value $J_c(N)$ for $\kappa = 0$
($CP^{N-1}$ model) is a decreasing function of $N$ and that
\cite{PV-19-CP,PV-20-largeNCP} $J_c(2) = 0.7102(1)$, $J_c(20)\approx
0.353$. Thus, the transitions we consider for $N\ge 2$ should belong
to the MH line. As a check, we have measured the Binder cumulant of
the bilinear operator $Q_{\bm x}$ defined in Eq.~(\ref{Qdef}). The
numerical results show that it converges to 1 on both sides of the
transitions as $L\to\infty$, confirming that the $SU(N)$ symmetry is
broken in both phases.

Note that $J=1$ is less than twice $J_c(N)$ at $\kappa=0$ for most of
the $N$ simulated ($N\le 10$), so we do not expect significant
crossover effects from the $IXY$ point at $J=\infty$.  This is not
true for $N=25$, so in this case we also performed simulations at
$J=0.4$ to verify that the results are independent of the specific
value of $J$ adopted. Close to the multicritical point where the three
transition lines meets, see Fig.~\ref{phdiasketchncLAH}, the
transitions along the MH line could in principle become discontinuous.
However, we have no indications of this behavior from the performed
simulations.

MC simulations have been performed by using a combination of
Metropolis and microcanonical updates, see, e.g.,
Ref.~\cite{BPV-21-ncAH} for more details.  To estimate mean values in
the Lorenz gauge, we have performed simulations using the weight
$e^{-H}$ (i.e. without gauge fixing) and implemented the gauge fixing
before each measure.  Given a MC configuration $\{A_{{\bm x},\mu},
{\bm z}_{\bm x}\}$, we have determined (by using a conjugate gradient
solver) the gauge transformation $\Lambda_{\bm x}\in\mathbb{R}$ such
that the new field $A'_{{\bm x},\mu}=A_{{\bm x},\mu} - \Lambda_{\bm
  x}+\Lambda_{{\bm x}+\hat{\mu}}$ satisfies the Lorenz condition
(\ref{Lgauge}).  Correlations are then computed using the
configuration $\{A'_{{\bm x},\mu}, {\bm z}'_{\bm x}\}$. It is simple
to show that this procedure is equivalent to directly sampling
configurations satisfying the Lorenz gauge condition with weight
$e^{-H}$.

\subsection{Critical points from the RG invariant observables}
\label{critpoints}

\begin{figure*}[!tbp]
\begin{tabular}{cc}
\includegraphics*[width=0.85\columnwidth]{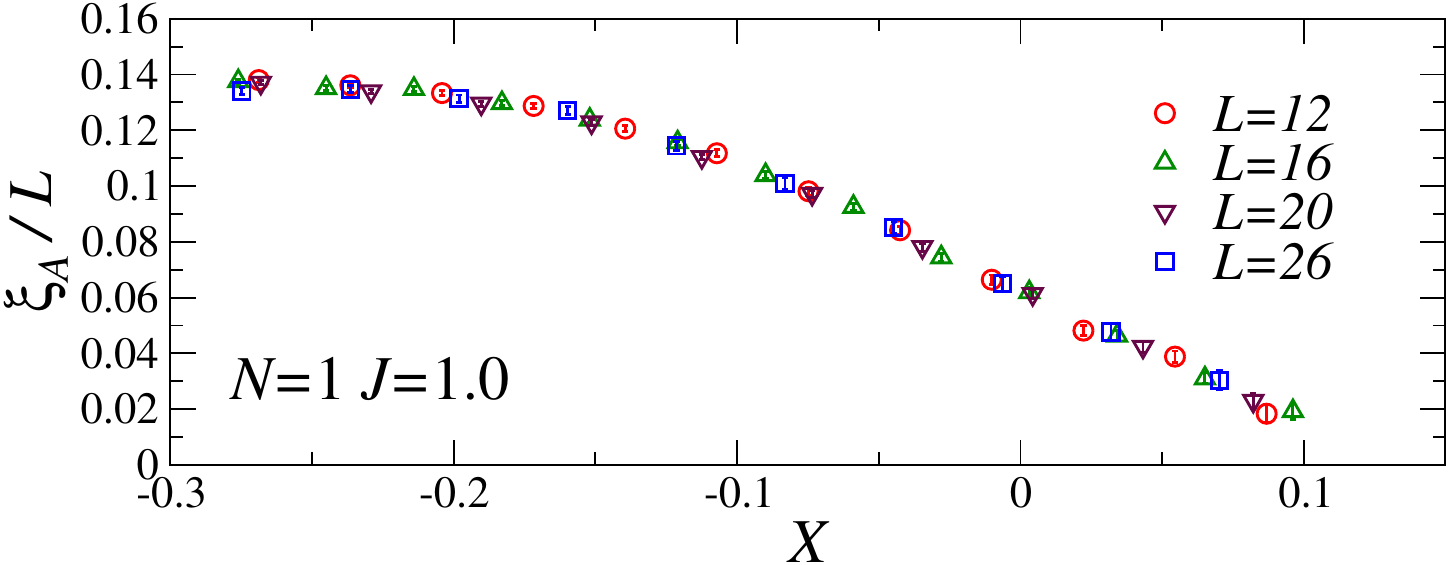} & 
\includegraphics*[width=0.85\columnwidth]{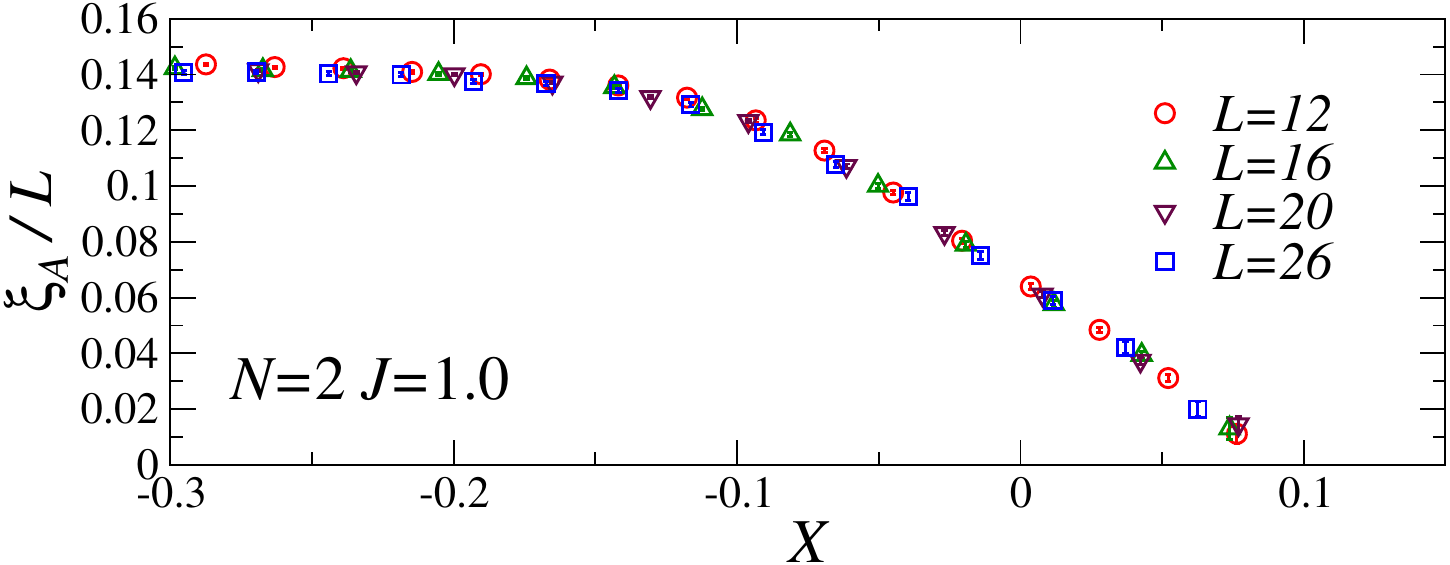} \\
\includegraphics*[width=0.85\columnwidth]{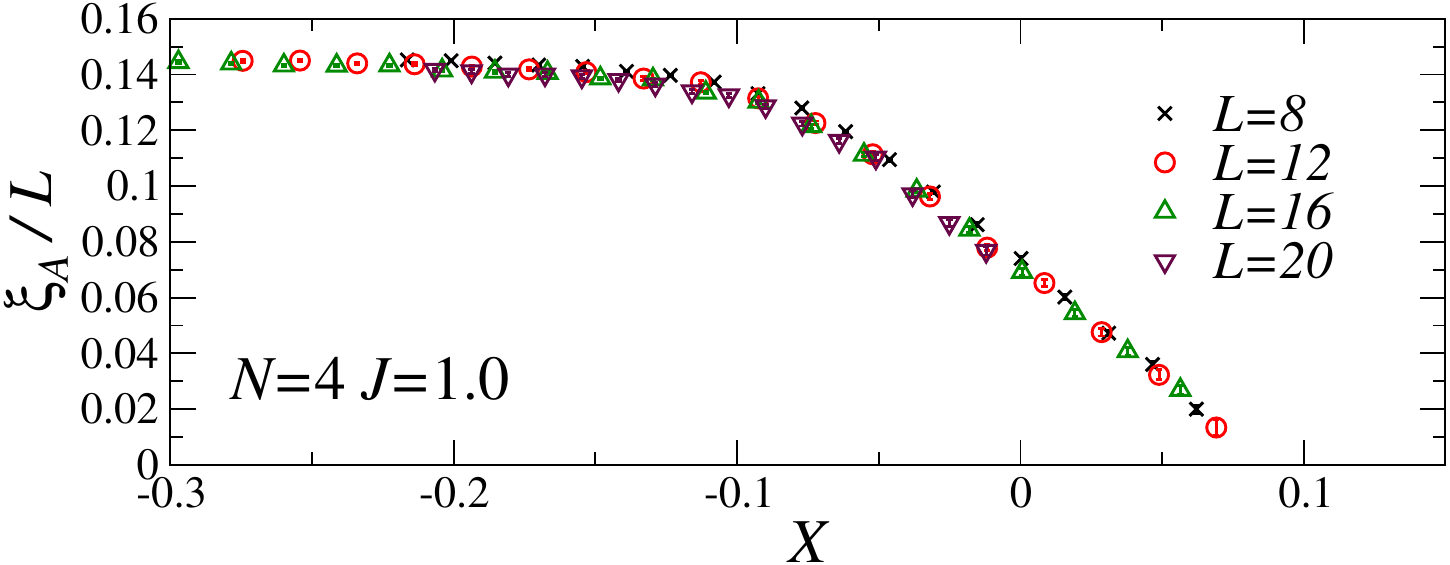} & 
\includegraphics*[width=0.85\columnwidth]{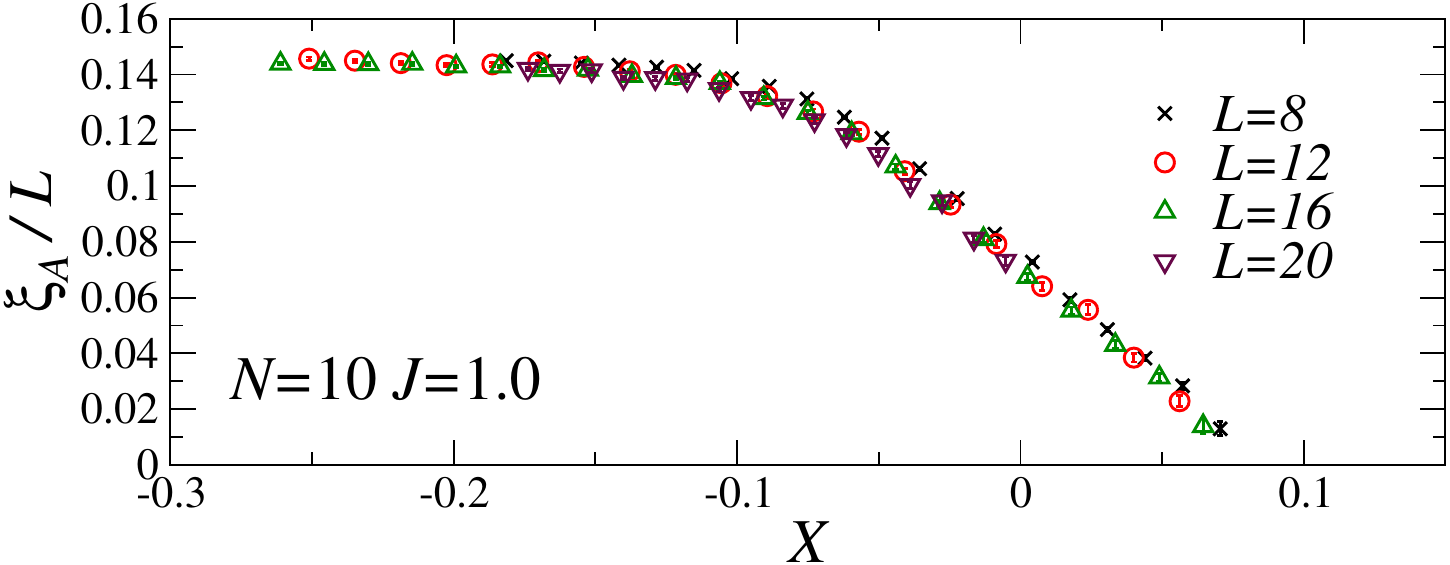} \\
\includegraphics*[width=0.85\columnwidth]{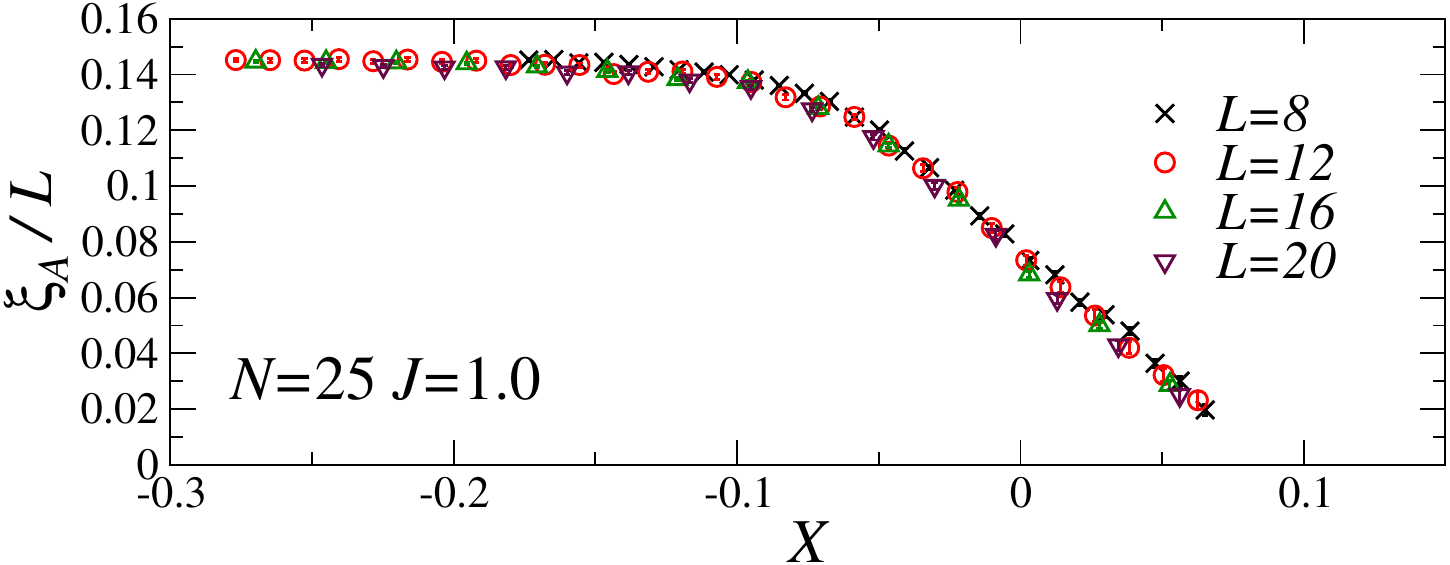} & 
\includegraphics*[width=0.85\columnwidth]{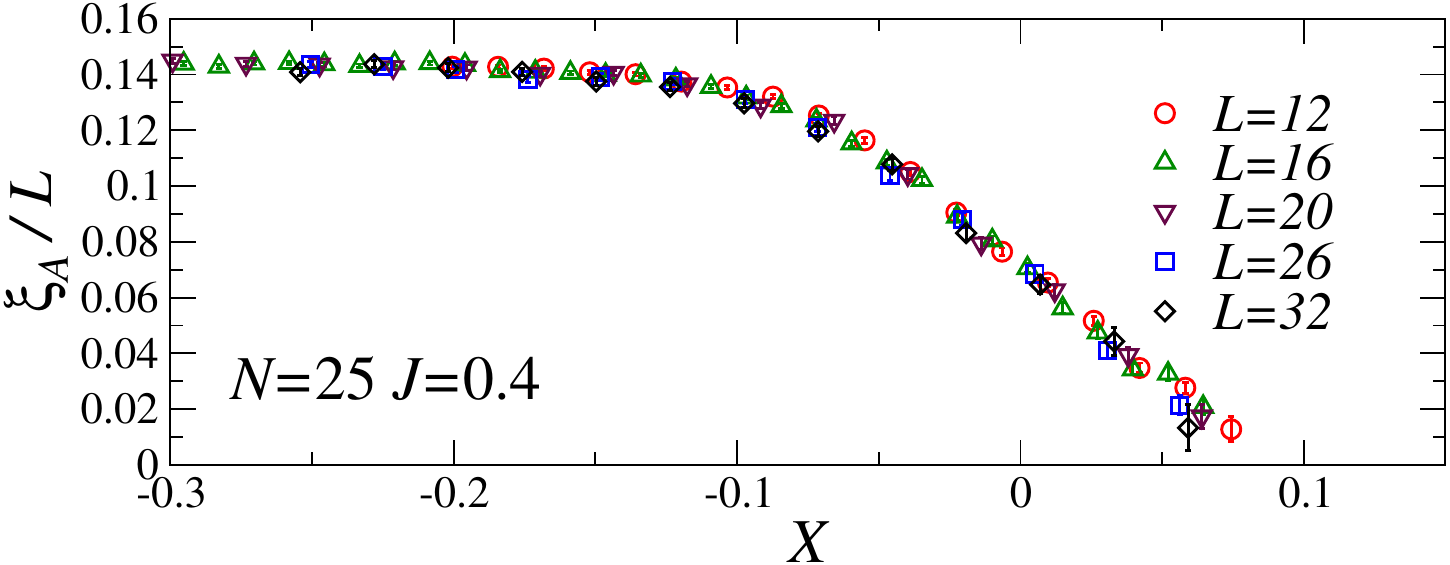} \\
\end{tabular}
\caption{Plot of $R_{A}=\xi_A/L$ versus $X = (\kappa - \kappa_c) L^{1/\nu}$
with $\nu = 0.6717$ for: 
a) $N=1$, $J=1$; 
b) $N=2$, $J=1$;
c) $N=4$, $J=1$;
d) $N=10$, $J=1$;
e) $N=25$, $J=1$; 
f) $N=25$, $J=0.4$.
}
\label{RxiAvsX} 
\end{figure*}

We have first determined the critical values $\kappa_c$, fitting the
MC data of $R_{A}$, $R_{z}$, $U_A$, and $U_z$ to Eq.~(\ref{eq:FSS1}).
For this purpose we have parameterized the corresponding scaling
functions ${\cal R}(X)$ with a polynomial in $X$ of degree $n$ (with
$n$ varying between 10 and 12). The fits allow us to determine both
$\kappa_c$ and $\nu$. Given the small range of available values of
$L$, we can only verify that the estimates of $\nu$ are substantially
consistent with the $XY$
value~\cite{CHPV-06,Hasenbusch-19,CLLPSSV-20,GZ-98,PV-02,KP-17},
$\nu_{XY}=0.6717(1)$. More evidence of a universal $IXY$ behavior is
provided below.  To obtain accurate estimates of $\kappa_c$ we have
repeated the fits fixing $\nu = 0.6717$.  We obtain
\begin{equation}\label{eq:critK}
\begin{array}{lll} 
N = 1\phantom{0}   & \  J = 1.0:    & \quad \kappa_c=0.10745(5);  \\
N = 2\phantom{0}   & \  J = 1.0:    & \quad \kappa_c=0.08931(5);  \\
N = 4\phantom{0}   & \  J = 1.0:    & \quad \kappa_c=0.08179(6);  \\
N = 10             & \  J = 1.0:    & \quad \kappa_c=0.07821(10); \\
N = 25             & \  J = 1.0:    & \quad \kappa_c=0.07685(8);  \\
N = 25             & \  J = 0.4:    & \quad \kappa_c=0.07996(4).  \\
\end{array}
\end{equation}
Note that for $J\to \infty$ the MH line is expected to converge to
$\kappa_{c,IXY} = 0.076051(2)$ for all values of $N$. Thus, the $J$
dependence of $\kappa_c$ becomes weaker as $N$ increases, in agreement
with the general argument presented in Sec.~\ref{largeN}.  The
scaling relation~(\ref{eq:FSS1}) is well satisfied by the data, with
scaling corrections that increase with increasing $N$ and decreasing
$J$. As an example in Fig.~\ref{RxiAvsX}, we report $R_{A}$ versus $X$
for the six different cases mentioned above.  The agreement is
excellent.  Moreover, in Table~\ref{table-Rstar} we report our
estimates $R_A^*$, $R_z^*$, $U_A^*$ and $U_z^*$ of the RG invariant
quantities $R_A$, $R_z$, $U_A$ and $U_z$ at the critical point of the
one-component and multicomponent LAH models that we have
considered. Their global agreement is a further robust evidence of
universality.

\begin{table}[t]
\caption{Estimates of the critical-point values of the RG-invariant
  quantities $R_z$, $R_A$, $U_z$ and $U_A$, for the different values
  of $N$. They turn out to be in good agreement, supporting
  universality.  }
\label{table-Rstar}
\begin{tabular}{lcccccc} 
  \hline\hline & $N=1$ & $N=2$ & $N=4$ & $N=10$ &
  \begin{minipage}{1.2cm} \rule{0mm}{3mm}$N=25$, \rule[-1mm]{0mm}{3mm}$J=1$
  \end{minipage} & 
  \begin{minipage}{1.2cm} \rule{0mm}{3mm}$N=25$, \rule[-1mm]{0mm}{3mm}$J=0.4$
  \end{minipage}\\ 
\hline
$R_{z}^*$ & 1.45(5) & 1.42(8) & 1.35(12) & 1.45(10) & 1.40(10) & $>1.25$ \\
$R_{A}^*$ & 0.064(4) & 0.067(5) & 0.075(9) & 0.069(6) & 0.073(6) & 0.071(6) \\
$U_{z}^*$ & 1.078(10)& 1.080(13)& 1.09(3) & 1.08(2) & 1.081(17) & 1.12(5) \\
$U_{A}^*$ & 3.3(1) & 3.5(1) & 3.6(2) & 3.7(2) & 3.7(3) & 3.2(2) \\
\hline\hline
\end{tabular}
\end{table}

\subsection{The gauge-invariant energy cumulants}
\label{enecumres}

\begin{figure}[tbp]
\includegraphics*[width=0.99\columnwidth]{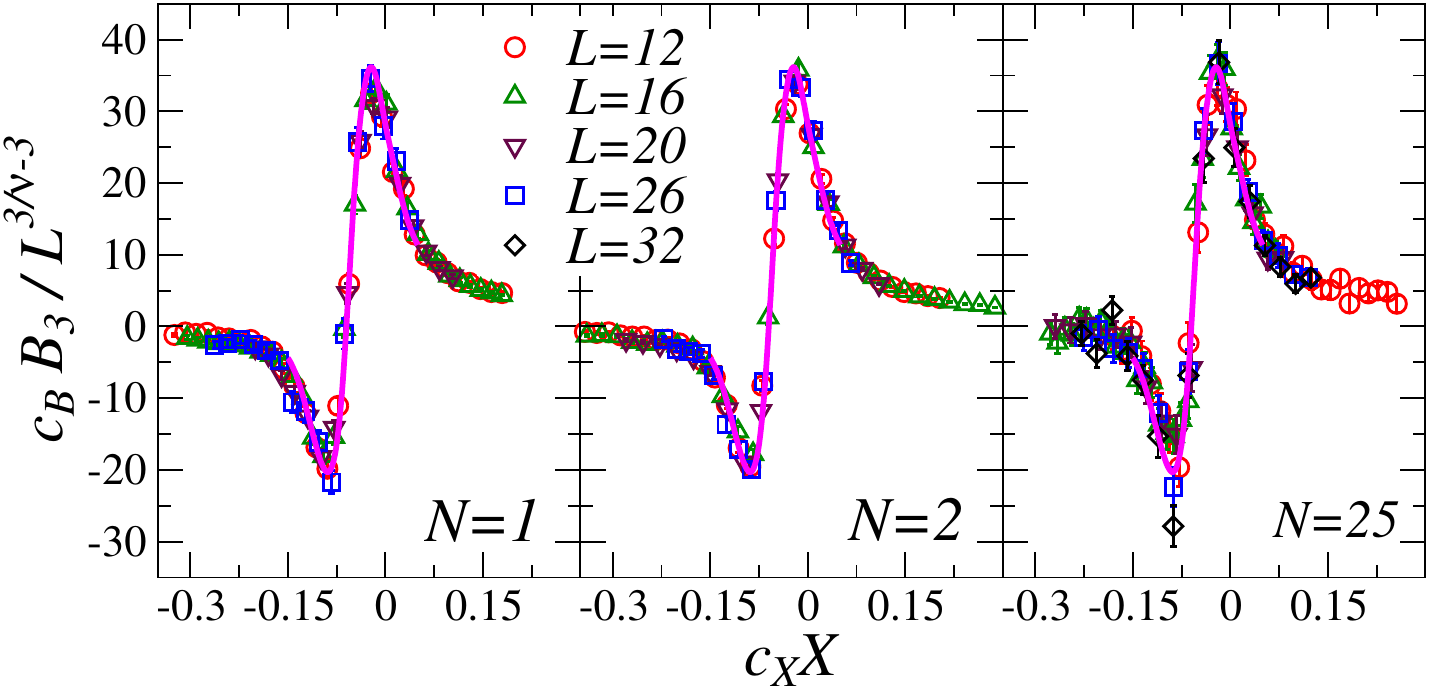}
\caption{Scaling plot of $B_3$ for $N=1$, $N=2$ (for $J=1$) and $N=25$
  (for $J=0.4$), from left to right.  We plot $\widetilde{\cal B}_3 =
  c_B B_3 L^{3-3/\nu}$ versus $\widetilde{X} = c_X(\kappa -
  \kappa_{c}) L^{1/\nu}$, fixing $\nu=\nu_{XY}\approx0.6717$ and the
  the critical values $\kappa_c=0.10745(5),\,0.08931(5),\,0.07996(4)$,
  respectively.  The continuous curve that appears in the three panels
  is the scaling curve computed in the $IXY$ model~\cite{BPV-24} with
  $c_X=c_B=1$. The nonuniversal constants $c_X$ and $c_B$ of the LAH
  models are fixed by matching the curves, obtaining $c_B =
  1.5,1.05,0.95$ and $c_X = 0.52,0.75,0.9$ for $N=1,2,25$,
  respectively. Analogous results are obtained for $N=4,\,10,\,25$
  along the $J=1$ line.}
\label{B3fig}
\end{figure}

MC results for $B_3$ are reported in Fig.~\ref{B3fig} for $N=1$,
$N=2$ (for $J=1$) and $N=25$ (for $J=0.4$). Analogous results are
obtained for $N=4,\,10,\,25$ at $J=1$.  The observed behaviors are
definitely consistent with Eq.~(\ref{Hg3-scaling}), when using the
$XY$ exponent $\nu_{XY}=0.6717$.  Note that, if we appropriately tune
the vertical and horizontal scale by adding two nonuniversal
constants, the scaling behavior is universal, i.e., the scaling curve
is quantitatively the same for all values of $N$ and for the $IXY$
model at $J=\infty$. The scaling curve of the $IXY$ model has been
computed in Ref.~\cite{BPV-24}, by performing an interpolation of MC
results for the $IXY$ model. Analogous results are obtained for the higher
cumulants.  We can therefore conclude that transitions along the MH line
for any $N\ge 2$ and along the CH line for $N=1$ have the 
same $IXY$ universal behavior.

\subsection{Nonlocal charged correlations}
\label{nonloccorr}

We now determine the critical behavior of the charged operator ${\bm
  \Gamma}_{\bm x}$ defined in Eq.~(\ref{phixdef}).  Charged
correlations are expected to have a nontrivial behavior along the MH
line.  Indeed, in the H phase we have $G_\Gamma({\bm x},{\bm y})\to c
\neq 0$ (thus, $G_z({\bm x},{\bm y})\to c \neq 0$ in the Lorenz gauge)
in the large $|{\bm x}-{\bm y}|$ limit, as demonstrated for
$N=1$~\cite{KK-85,KK-86,BN-86,BN-86-b} and numerically confirmed for
multicomponent systems~\cite{BPV-23b}.

\begin{figure*}[!btp]
\begin{tabular}{cc}
\includegraphics*[width=0.85\columnwidth]{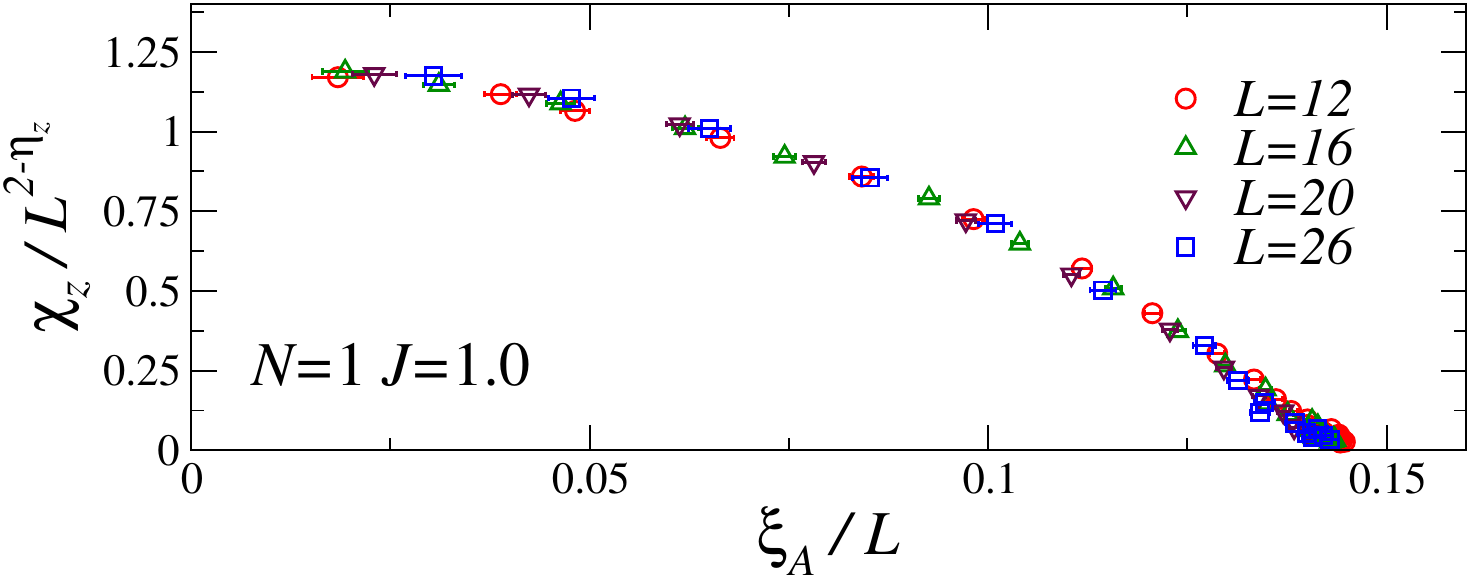} & 
\includegraphics*[width=0.85\columnwidth]{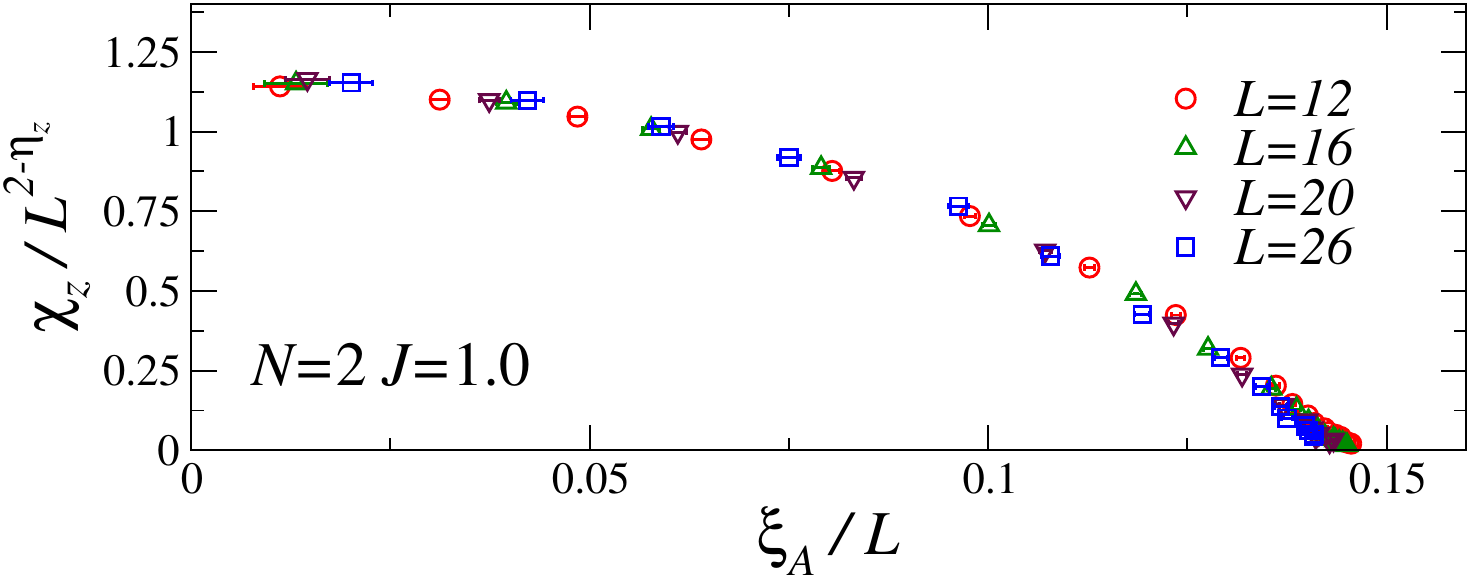} \\
\includegraphics*[width=0.85\columnwidth]{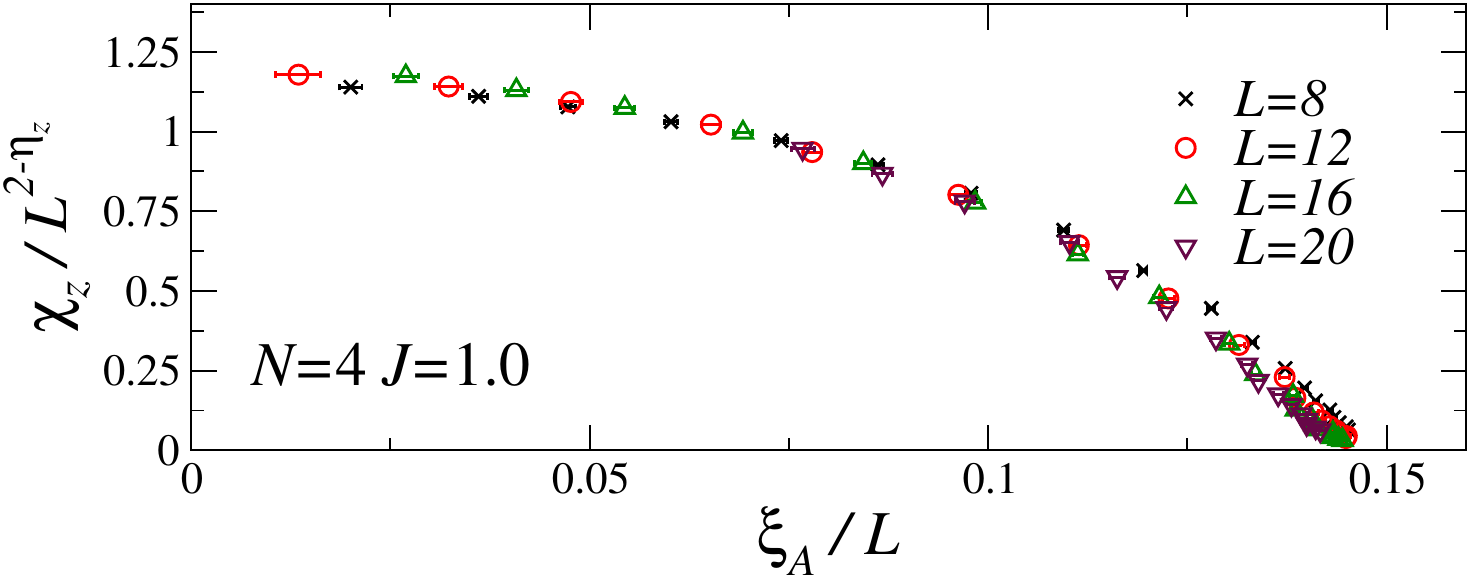} & 
\includegraphics*[width=0.85\columnwidth]{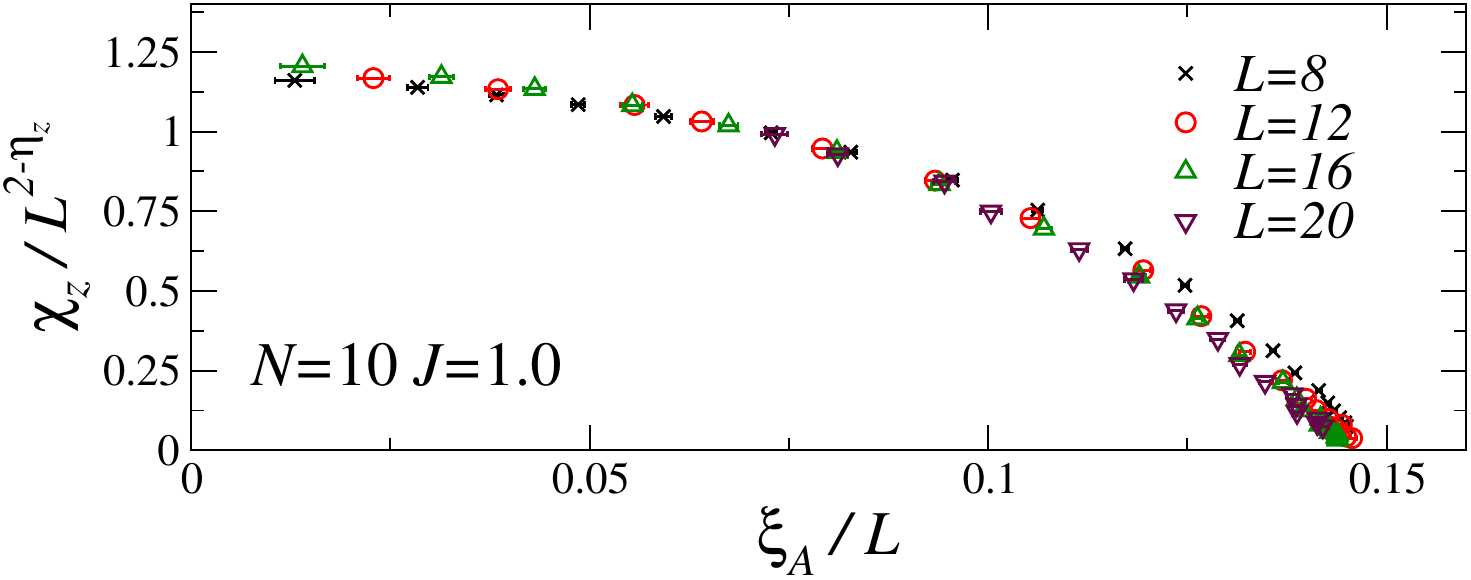} \\
\includegraphics*[width=0.85\columnwidth]{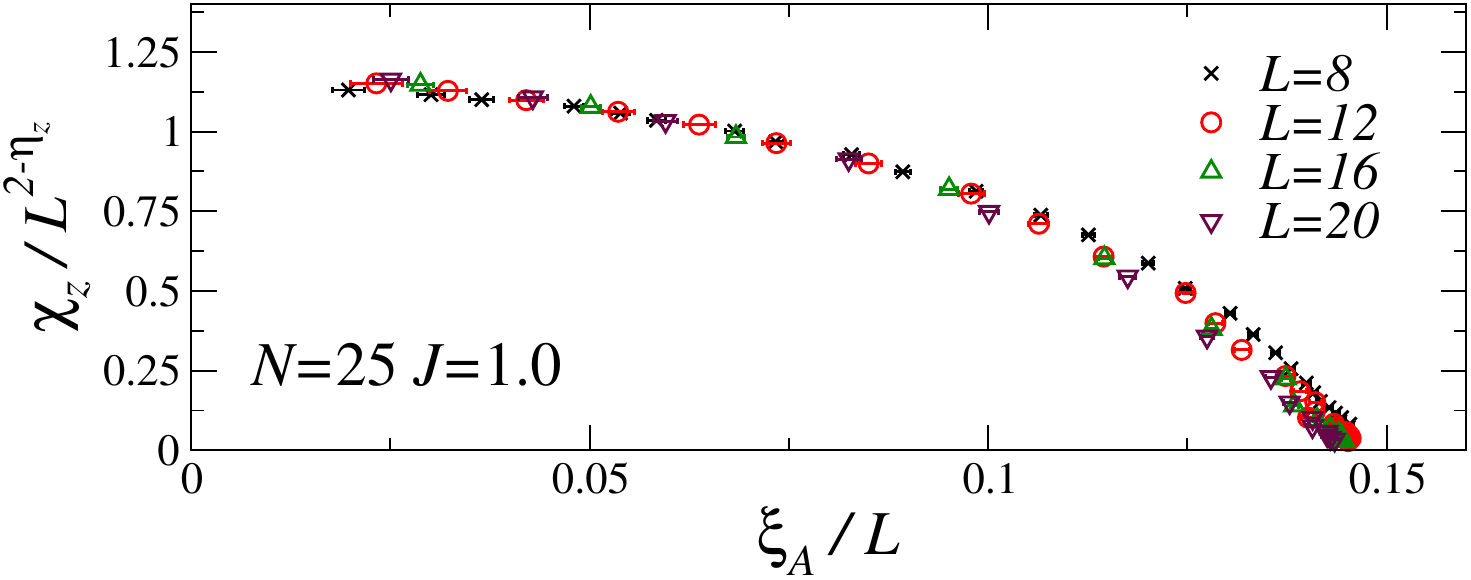} & 
\includegraphics*[width=0.85\columnwidth]{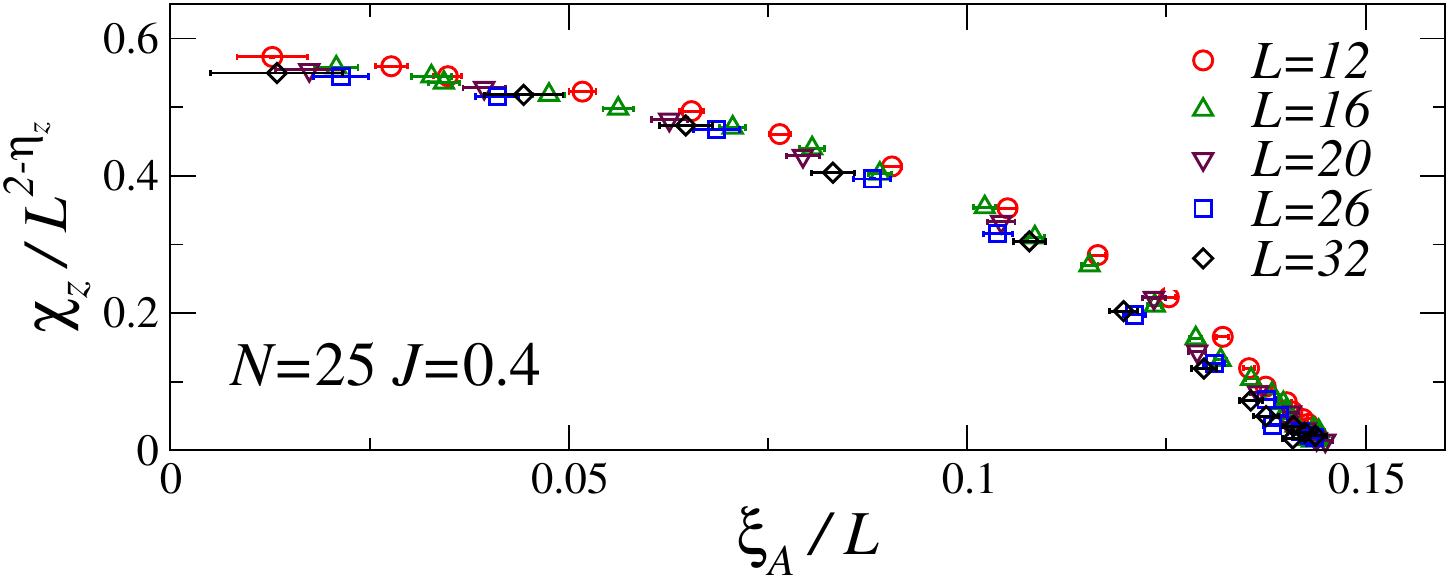} \\
\end{tabular}
\caption{Plot of $\chi_z/L^{2-\eta_z}$ for $\eta_z = -0.74$ versus
  $R_{A}=\xi_A/L$ for: a) $N=1$, $J=1$; b) $N=2$, $J=1$; c) $N=4$, $J=1$; d)
  $N=10$, $J=1$; e) $N=25$, $J=1$; f) $N=25$, $J=0.4$.  The data
  approach a unique curve, apart from a multiplicative normalization,
  supporting universality.}
\label{chizvsRxiA} 
\end{figure*}

The numerical results confirm that charged correlations have an
$N$-independent critical behavior, which is the same as that occurring
in the one-component LAH model along the CH line.  In particular, the
Lorentz-gauge susceptibility $\chi_z$ scales as
\begin{equation}
  \chi_z\approx L^{2-\eta_z} F_\chi(X),
  \label{chizsca}
  \end{equation}
with $\eta_z$ independent of $N$.  The analysis of the scalar
correlations in the Lorenz gauge gives $\eta_z = -0.74(4)$, 
$-0.76(2)$,  $-0.74(3)$, $-0.75(3)$, $-0.72(4)$ for 
$N=1,2,4,10,25$ (in all cases $J = 1$). We thus end up with 
the $N$-independent estimate
\begin{equation}
  \eta_z = -0.74(4).
  \label{etazest}
\end{equation}
The plots of the MC data of $\chi_z$ shown in Fig.~\ref{chizvsRxiA}
nicely support the scaling behavior (\ref{chizsca}) with the above
estimate of $\eta_z$.

\begin{figure}[tbp]
\includegraphics*[width=0.99\columnwidth]{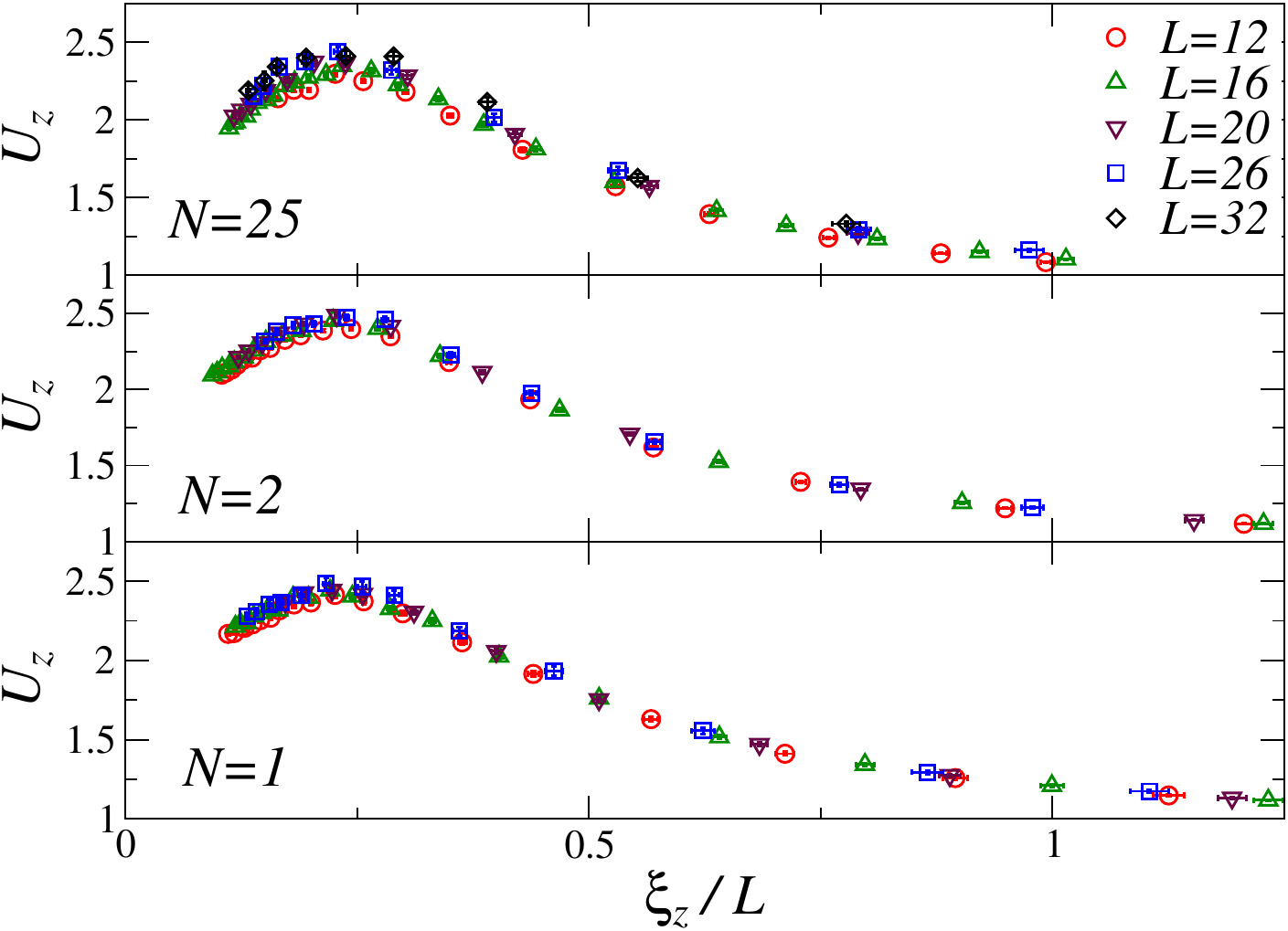}
\caption{The Binder parameter $U_z$ versus the ratio $R_z=\xi_z/L$ in
  the Lorenz gauge, for $N=25$ with $J=0.4$ (top), $N=2$ (middle) and
  $N=1$ (bottom) with $J=1$.  They appear to converge to the same
  universal curve.  Analogous results are obtained for $N=4,\,10,\,25$
  at $J=1$.}
\label{UzRz}
\end{figure}

To provide additional, and more compelling, evidence that the critical
MH transitions belong to the same universality class, irrespective of
the value of $N$, we consider the FSS relation $U_z\approx F_U(R_z)$
between the Binder parameter and $R_z\equiv \xi_z/L$.  For given
boundary conditions and lattice shape, the function $F_U$ depends only
on the universality class, without requiring the tuning of
nonuniversal parameters~\cite{PV-02,BPV-21-ncAH}.  The scaling curves
shown in Fig.~\ref{UzRz} are the same for all values of $N$,
confirming that the critical behavior of the charged sector is $N$
independent.

It is interesting to observe that these results are also relevant for the 
$IXY$ model as they characterize the critical behavior of the nonlocal 
operator $\widetilde{\Gamma}_{\bm x}$, that corresponds to the insertion 
of static charges in the system.

\subsection{Local gauge correlations}
\label{gaucorr}

Gauge and charged correlations show a critical behavior for any
$N$. The FSS analysis of the $F_{{\bm x},\mu\nu}$ correlations or,
equivalently, of the susceptibility $\chi_A=V^{-1}\sum_{{\bm x}{\bm
    y}} C_{\mu\mu}({\bm x},{\bm y})$ in the Lorenz
gauge~\cite{BPV-23b}, allows us to estimate the gauge-field exponent
$\eta_A$ ($\chi_A \sim L^{2-\eta_A}$ at the critical point).  In the
$IXY$ model we have $\eta_A = 1$, an exact result that follows from
the correspondence between the small-momentum correlation of $F_{{\bm
    x},\mu\nu}$ in the $IXY$ model and the helicity modulus $\Upsilon$
computed in the dual $XY$ model~\cite{NRR-03}, and from the fact that
$\Upsilon$ scales as $L^{-1}$ in the $XY$ model~\cite{FBJ-73}.  If the
critical behavior along the MH line (CH line for $N=1$) is the same as
in the $IXY$ model, we expect $\eta_A = 1$ in all cases.

This result is confirmed by the numerical analyses of $\chi_A$ for all
values of $N$ considered.  Note that $\eta_A = 1$ also holds along the
CH line for $N> N^{\star}$, where the transition is controlled by the
charged fixed point of the RG flow of the
AHFT~\cite{BPV-23b,HT-96,HS-00,KNS-02,BFLLW-96}, and, more generally,
at any continuous transition controlled by a charged field-theoretical
fixed point, as a consequence of the Ward identities~\cite{ZJ-book}.

\begin{figure*}[!htbp]
\begin{tabular}{cc}
\includegraphics*[width=0.85\columnwidth]{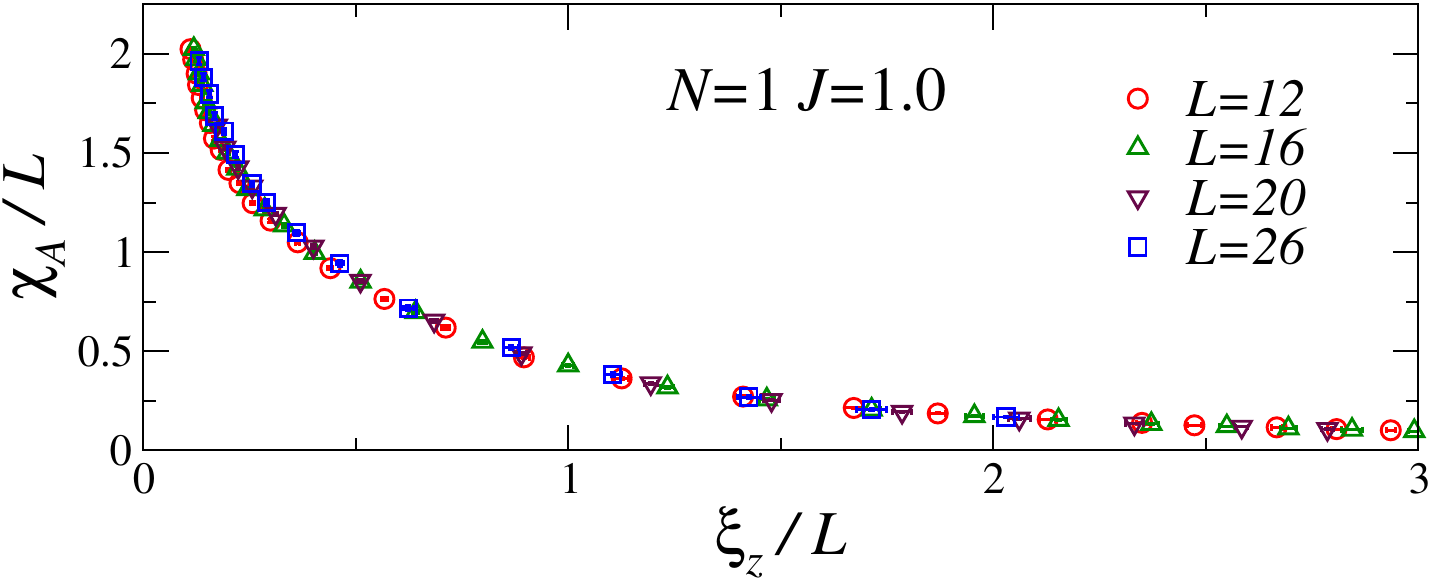} & 
\includegraphics*[width=0.85\columnwidth]{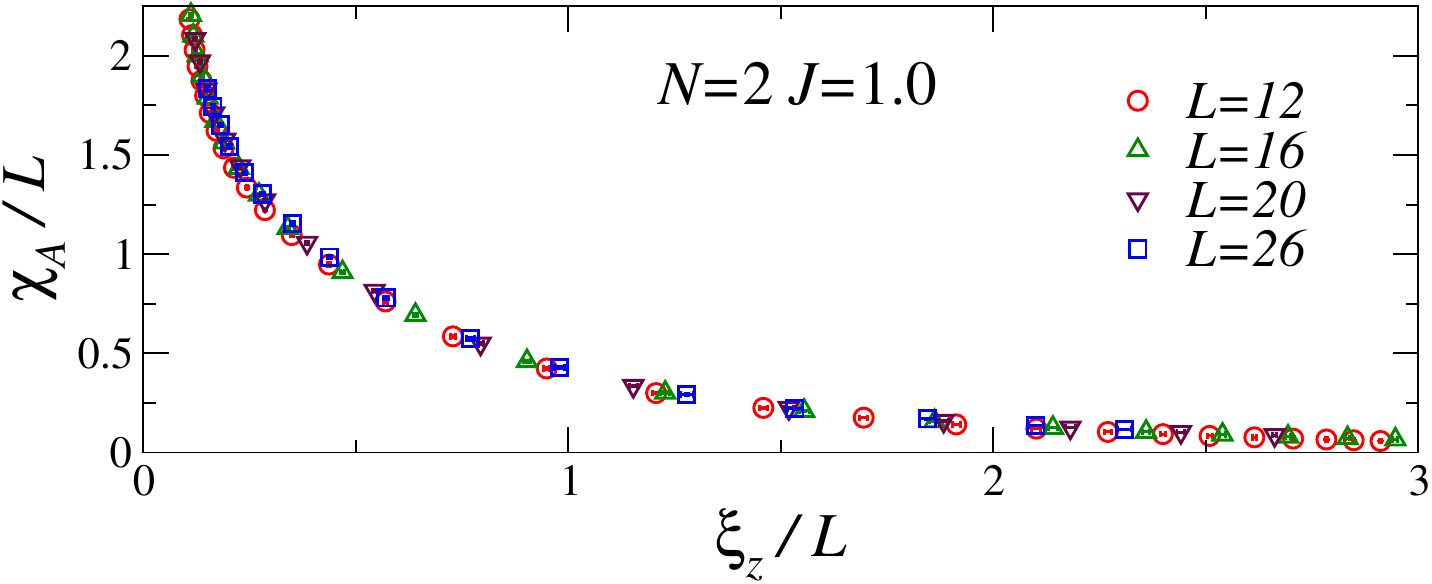} \\
\includegraphics*[width=0.85\columnwidth]{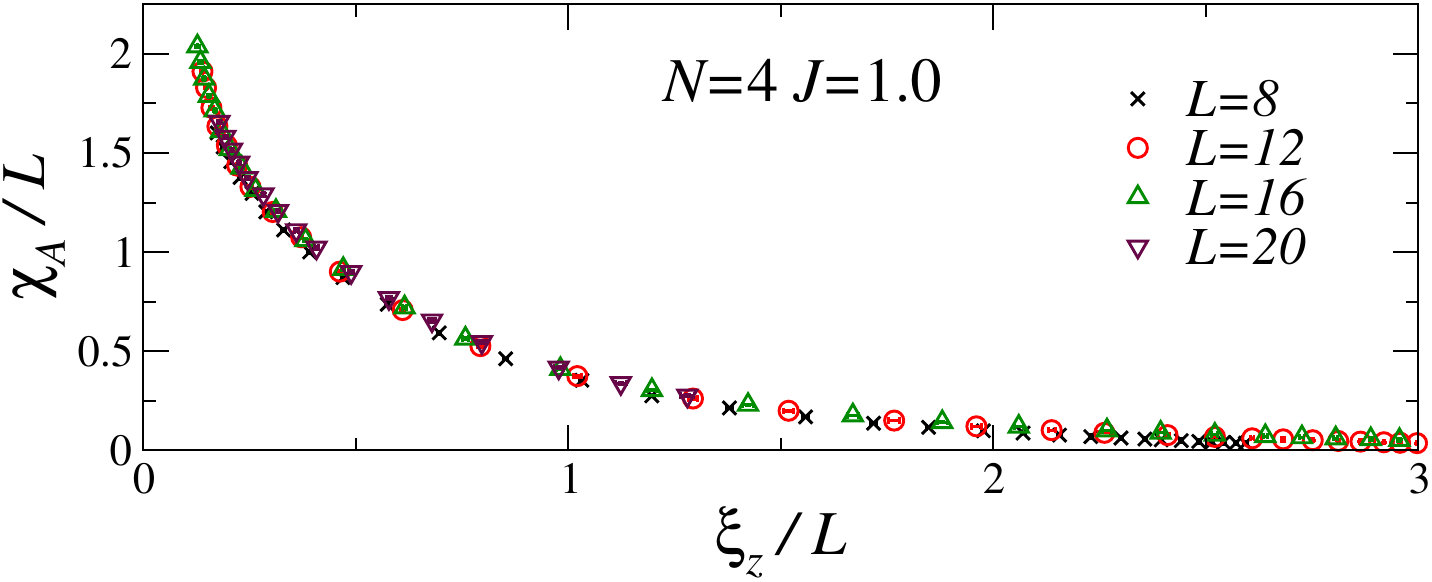} & 
\includegraphics*[width=0.85\columnwidth]{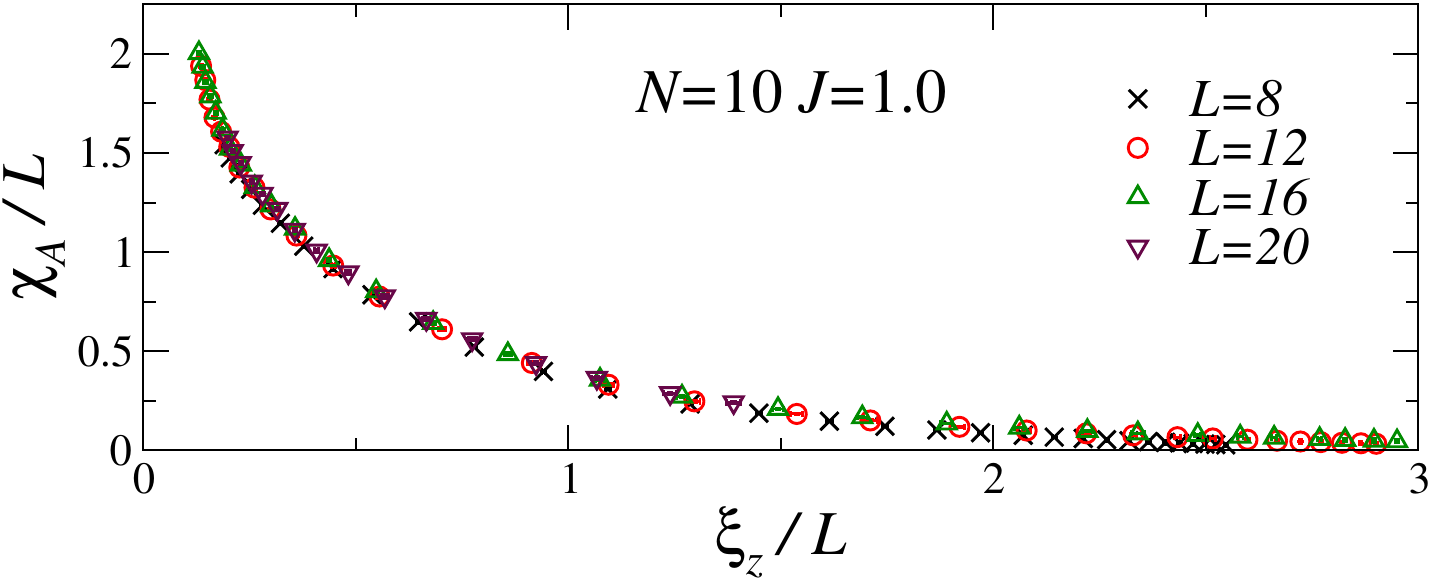} \\
\includegraphics*[width=0.85\columnwidth]{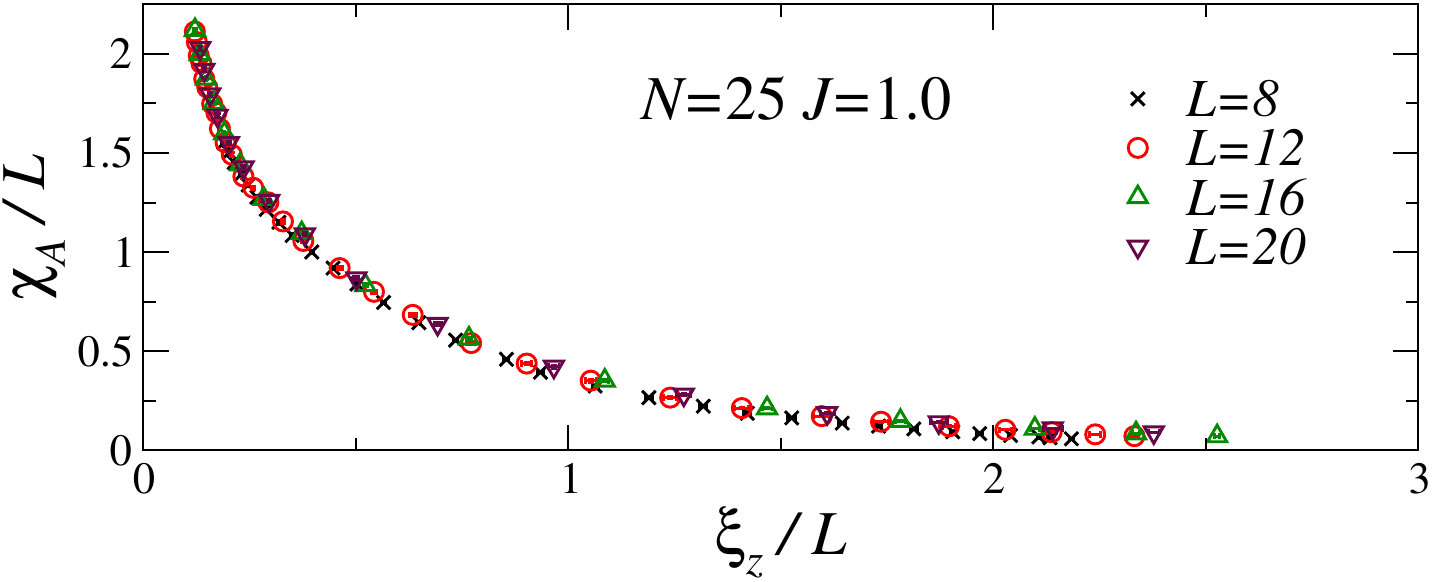} & 
\includegraphics*[width=0.85\columnwidth]{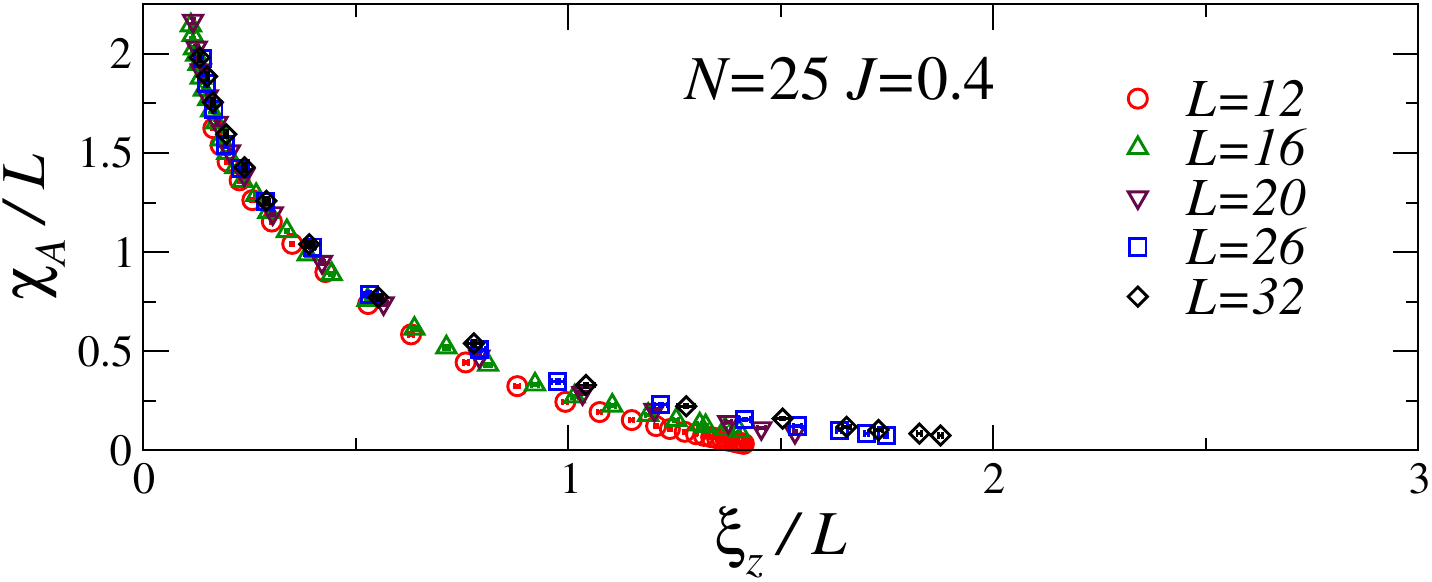} \\
\end{tabular}
\caption{Plot of $\chi_A/L^{2-\eta_A} = \chi_A/L$ ($\eta_A = 1$)
  versus $R_{z}=\xi_z/L$ for: a) $N=1$, $J=1$; b) $N=2$, $J=1$; c) $N=4$,
  $J=1$; d) $N=10$, $J=1$; e) $N=25$, $J=1$; f) $N=25$, $J=0.4$.  The
  data approach a unique curve, apart from a multiplicative
  normalization, supporting universality.  For $R_{z}\to 0$ data scale
  as $1/R_{z}$, to guarantee the correct Coulomb behavior of the
  susceptibility $\chi_A \sim L^2$. }
\label{chiAvsRxiz} 
\end{figure*}

\section{Conclusions}
\label{conclu}

We have studied the topological phase transitions occurring in the 3D
LAH model, in which an $N$-component scalar field is minimally coupled
with a noncompact Abelian gauge field, with a global $SU(N)$
symmetry. The phase diagram, see Fig.~\ref{phdiasketchncLAH}, presents
three phases, which differ in the properties of the gauge
correlations, in the confinement or deconfinement of the charged
excitations, and in the behavior under global $SU(N)$ transformations.

We have shown that multicomponent LAH models can undergo different
types of transitions driven by the condensation of charged
excitations: CH and MH transitions along the lines that separate the H
phase from the C and M phase, respectively. They are controlled by
different {\em charged} fixed points of the renormalization-group (RG)
flow with nonvanishing gauge coupling.  Gauge correlations play an
active, but different, role at these deconfinement transitions and,
therefore, they cannot be described by effective LGW theories.  While
transitions along the CM and CH lines are related to the spontaneous
breaking of the global $SU(N)$ symmetry, the MH line separates two
ordered phases, both characterized by the condensation of a
gauge-invariant bilinear of the scalar field. The MH transitions are
driven by gauge modes that undergo a critical transition without a
local gauge-invariant order parameter, as it also occurs for the CH
transitions in the one-component LAH model, see
Fig.~\ref{phdiasketchncLAH}.

The numerical results we have presented show that, for any $N\ge 2$
(including the $N\to\infty$ limit), the topological MH transitions
belong to the same universality class as the transitions along the CH
line for $N=1$.  Charged excitations associated with ${\bm
  \Gamma}_{\bm x}$, deconfine at the transition.  In the Lorenz gauge,
${\bm \Gamma}_{\bm x}$ is equivalent to the scalar field ${\bm z}_{\bm
  x}$ and it is therefore a local operator, allowing us to exploit the
standard RG framework.  We estimate the RG dimension of ${\bm
  \Gamma}_{\bm x}$, obtaining $y_\Gamma = (d-2+\eta_z)/2 \approx
0.13(2)$. We observe that excitations with RG dimensions $y_\Gamma$
also exist in the $IXY$ model, i.e., in the absence of explicit scalar
fields, and are associated with the operator $\widetilde{\Gamma}_{\bm
  x}$ defined in Eq.~(\ref{phixdef-IXY}).  We finally note that
charged excitations are also relevant along the CH
line~\cite{BPV-23b}.  The ${\bm \Gamma}_{\bm x}$ operator has here a
different $N$-dependent RG dimension $y_\Gamma$: $y_\Gamma =
0.4655(5)$ for $N=25$~\cite{BPV-23b} and $y_\Gamma \approx 1/2
-10/(\pi^2 N)$ for large $N$~\cite{HLM-74,KS-08}.

We have also discussed the critical behavior of gauge local
correlations, showing that the associated susceptibility exponent
$\eta_A$ satisfies $\eta_A=1$, as in the $IXY$ model. For the latter
model $\eta_A=1$ is an exact result that can be proved using duality.

Note that, although the $IXY$ model is dual to the $XY$ model, the
critical behavior of the charged scalar correlations along the $N = 1$
CH line is not related to that of the scalar correlations in the $XY$
model obtained for $\kappa\to\infty$, see Fig.~\ref{phdiasketchncLAH}.
Obviously, the Lorenz-gauge correlations of the field ${\bm z}_{\bm
  x}$ along the CH line converge to the scalar correlations of the
$XY$ model for $\kappa\to\infty$ at fixed system size $L$. However,
since the $\kappa\to\infty$ and the $L\to \infty$ limits do not
commute, this result does not imply that their asymptotic
infinite-volume behavior is the same.  Indeed, charged scalar
correlations are characterized by the critical exponent $\eta_z\approx
-0.7$ along the CH line at finite $\kappa$, definitely differing from
the value $\eta_z\approx 0.038$ at the $\kappa=\infty$ $XY$
transition~\cite{PV-02}. This is consistent with the RG field-theory
result that predicts the $XY$ RG fixed point to be unstable with
respect to gauge fluctuations~\cite{BPV-21-ncAH}.

Although the CH and MH transitions both separate a phase with confined
charges from a deconfined Higgs phase, their critical behavior shows
notable differences, since the transitions are associated with
different charged RG fixed points.  Indeed, the critical MH
transitions are effectively controlled by the charged $IXY$ fixed
point, while the continuous CH transitions, which occur for $N >
N^\star\approx 7$~\cite{IZMHS-19,BPV-21-ncAH,BPV-22,SZJSM-23}, are
controlled by the different, $N$-dependent, charged fixed point that
is obtained in the AHFT.  Estimates of the corresponding $N$-dependent
critical exponents are reported in Ref.~\cite{BPV-21-ncAH,BPV-23b};
for instance, $\nu=0.802(8)$ for $N=25$, and $\nu\approx 1 - 48/(\pi^2
N)$ for large $N$~\cite{HLM-74,MZ-03}. Therefore, LAH models with
$N>N^\star$ show two distinct charged critical behaviors along the MH
and CH transition lines.  Note that, if the MH and CH transitions
remain continuous up to their intersection point (they could turn into
first-order transitions before it), a nontrivial multicritical
behavior can occur~\cite{PV-02}, calling for further investigations.
We finally remark that the $N$-independent MH critical behavior does
not have a counterpart in the RG flow of the AHFT, as determined using
the perturbative $\epsilon=4-d$ expansion~\cite{HLM-74,IZMHS-19}, or
in the standard large-$N$ approach~\cite{MZ-03}.

It would be interesting to add fermionic fields to the LAH model,
minimally coupled to the gauge field.  Also in this case one expects
Higgs phases bounded by topological transitions.  In particular, for
large $J$ one still expects a transition line where gauge fields
behave as in the $IXY$ model.  While massive fermions should not
change the critical behaviors (they can be effectively integrated
out), massless fermions may lead to different topological universality
classes, which may be of interest for condensed-matter systems. At
these transitions the charged gauge-invariant fermionic
excitations~\cite{Dirac:1955uv} may be studied as the scalar
ones. Topological charged transitions driven by gauge fields can also
be present in 3D non-Abelian gauge theories with scalar fields in
diverse representations, see, e.g.,
Refs.~\cite{Sachdev-19,SSST-19,BPV-19,SPSS-20,BFPV-21}.  However, in
this case the identification of the relevant gauge-invariant matter
excitations is more complex.  In particular, the nonlinearity of the
gauge conditions~\cite{Faddeev:1967fc,Gribov:1977wm,Singer:1978dk}
does not allow us to extend to the non-Abelian case the simple
gauge-fixing approach exploited in this work. These issues call for
further investigation.

\end{document}